\documentclass[letterpaper,twocolumn,10pt]{article}
\usepackage{usenix2019_v3}
\pdfoutput=1
\usepackage{breakurl}           
\usepackage{url}                
\usepackage{tabularx}
\usepackage{booktabs}
\usepackage{amsmath,amssymb,amsfonts}
\usepackage{algorithmic}
\usepackage{float}
\usepackage{graphicx}
\usepackage{textcomp}
\usepackage{xcolor}
\usepackage{xspace}
\usepackage{graphicx}
\usepackage{subfig}
\def\BibTeX{{\rm B\kern-.05em{\sc i\kern-.025em b}\kern-.08em
    T\kern-.1667em\lower.7ex\hbox{E}\kern-.125emX}}
    
\makeatletter
\let\c@subfigure\relax
\let\c@subtable\relax
\let\@listsubcaptions\relax
\let\@dottedxxxline\relax
\let\l@subfigure\relax
\let\c@lofdepth\relax
\let\l@subtable\relax
\let\c@lotdepth\relax

\let\subfloat@label\relax
\let\sf@@sub@label\relax

\makeatother
\usepackage[tight]{subfigure}
 
\newcommand{\paraspace}{\vspace{0.03in}}
\newcommand{\parab}[1]{\paraspace\noindent{\bf #1}}
\newcommand{\sys}{{ESA}\xspace}
\newcommand{\sysname}{{ESA}\xspace}

\newcommand{\TODO}[1]{{\color{red}{[TODO: #1]}}}

\newcommand{\yanfang}[1]{\textbf{\color{blue}yle: #1}}
\newcommand{\lam}[1]{\textbf{\color{orange}lam: #1}}
\newcommand{\wh}[1]{\textbf{\color{brown}wh: #1}}
\begin{document}

    \title{Efficient Data-Plane Memory Scheduling for In-Network Aggregation}
    %

    \author{Hao Wang$^1$, Yuxuan Qin$^1$, ChonLam Lao$^2$, Yanfang Le$^3$, Wenfei Wu$^4$, Kai Chen$^1$\\$^1$iSING Lab, Hong Kong University of Science and Technology\\
$^2$Harvard University $^3$Intel, Barefoot Switch Division $^4$Peking University}
    \maketitle
    \begin{abstract}

As the scale of distributed training grows, communication becomes a bottleneck. To accelerate the communication, recent works introduce In-Network Aggregation (INA), which moves the gradients summation into network middle-boxes, e.g., programmable switches to reduce the traffic volume.
However, switch memory is scarce compared to the volume of gradients transmitted in distributed training. Although literature applies methods like pool-based streaming or dynamic sharing to tackle the mismatch, switch memory is still a potential performance bottleneck. Furthermore, we observe the under-utilization of switch memory due to the synchronization requirement for aggregator deallocation in recent works.


To improve the switch memory utilization, we propose \sys, an \underline{E}fficient Switch Memory \underline{S}cheduler for In-Network \underline{A}ggregation. At its cores, \sys enforces the preemptive aggregator allocation primitive and introduces priority scheduling at the data-plane, which improves the switch memory utilization and average job completion time (JCT). Experiments show that \sys can improve the average JCT by up to $1.35\times$. 

\end{abstract}

    \vspace{-0.1in}
\section{Introduction}
\vspace{-0.1in}


Driven by the increasing complexity of Machine Learning (ML) applications, such as autonomous driving~\cite{levinson2011towards}, general-purpose language models~\cite{wang2019superglue}, and game AI~\cite{wang2016does}, the ML models' size is increasing explosively, from ResNet50~\cite{resnet2016cvpr} with 23M parameters to GPT-3~\cite{gpt3} with 175B parameters. ML practitioners usually leverage distributed training (DT) systems to parallelize the modeling training algorithms for large models and datasets, and communication could become the bottleneck in this case.

Recent works propose In-Network Aggregation (INA)~\cite{switchml, atp, iswitch, panama, hire} to overcome this bottleneck, but also elicit a \emph{new critical resource requirement: the switch memory}. INA solutions organize \emph{switch memory}, i.e., SRAM and TCAM, into \emph{aggregators}, and place the gradients reduction into aggregators; such solutions can significantly reduce the in-network traffic volume, eliminate the bottleneck, and consequently accelerate the whole training job (e.g., 5.5X acceleration in SwitchML~\cite{switchml} for 8-server training DeepLight~\cite{legendre2019deeplight}). Allocating a DT job with sufficient switch memory could effectively promote its training speed, otherwise, the DT job has to fall back to the original communication mode without such performance gain.

However, switch memory is a scarce resource to support multiple INA jobs. Typical physical programmable switches,  such as Tofino~\cite{p4mem}, have only about 10MB memory per pipeline. And the physical memory is not used only for INA in practice --- it also needs to support other network functionalities such as load balancing~\cite{netcache}. The remaining memory can hardly support massive INA jobs. For example, one single job in SwitchML~\cite{switchml} takes up 1 MB in a 100Gbps setting; however, a production environment could have more than 5000 GPUs and tens of thousands of daily jobs~\cite{antman}. Thus, devising a reasonable switch memory usage model could efficiently promote massive DT jobs' performance. 

Existing INA solutions have not utilized switch aggregators to the best extent. One class of INA solutions~\cite{switchml, iswitch, panama} statically partition the switch memory into isolated regions, and each job is assigned with one region. The switch memory is occupied at the granularity of a job's lifetime, which is not efficiently used when the job is in computation. Another class~\cite{atp} maintain the switch memory as a shared resource pool and serve each job's each \emph{aggregation task} (defined as aggregating packets of the same sequence number from all workers) with non-preemptive first-come-first-serve (FCFS). The switch memory is occupied at the granularity of each aggregation task, which could be easily disturbed by the synchronization delay and worker stragglers. 

We propose \sys to efficiently utilize the switch aggregators. \sys is like a ``cache replacement'' system: once an aggregator (a cache unit) is not accessed for a while, it becomes cold and is replaced by another hot item. \sys consists of a data plane and a control plane. In the data plane, \sys has a priority-based preemption mechanism, where a task (packets) with a high priority can evict an existing low-priority task at the aggregator. In the control plane, \sys makes three intuitions to assign priority to tasks: model front layers, models with high communication-to-computation ratio, and models with a shorter remaining time should have higher priority. 

In the design of \sysname, we overcome three challenges. 
First, in most existing INA solution implementations, the complex aggregation logic has already used up the switch resources, e.g., ATP exhausts all meter ALUs of stages 4-10 (total 12 stages), it is challenging to implement an extra preemption mechanism. \sysname uses end-host PS to assist the switch aggregation, which handles various complex but low-probability corner cases, i.e., preemption or packet loss.
Second, preemption complicates the correctness guarantee at the transport layer. In a preemption, the evicted task's packets are separated as two parts, and the \emph{partial result packets} are more likely to lead to memory leak, wasted computation, and sliding window blocking. \sysname devises a reliability channel and a reminder mechanism to handle \emph{all-case} failures. 
Third, preemption should be carefully tuned. Low-frequency preemption would cause aggregators to spend more time in idle, and high-frequency preemption would cause aggregators to spend time on thrashing instead of aggregating. To avoid the idle or thrashing at aggregator (and the consequent performance degradation), \sysname devises a priority assignment and \emph{downgrading} policy for aggregation tasks.

We have implemented an \sys prototype with programmable switches. We evaluate \sys on a small testbed with one Wedge100 programmable switch and 10 NVIDIA V100s and a 64-node NS3~\cite{ns3} simulation as a complement. In our experiment, we find that \sys outperforms SwitchML and ATP by up to 1.89$\times$ and 1.35$\times$. Our deep dives further validate the effectiveness of our two main ideas, i.e., preemptive allocation and priority scheduling.

    \vspace{-0.1in}
\section{Background \& Motivation}
\label{sec:background}
\vspace{-0.1in}

\subsection{Preliminaries}
\vspace{-0.1in}

\parab{Distributed Training.} In various existing ML applications ~\cite{resnet2016cvpr,chowdhury2003natural,hoy2018alexa,levinson2011towards}, the required computational resources notably exceed a single machine's capability (>1000 Pflops-day v.s. <2.5 Pflops), and the modeling training becomes distributed. Data parallelism is one of the most common approaches in distributed training~\cite{bytescheduler,poseidon}. The distributed training algorithm is iterative, and each iteration consists of three steps: (1) forward propagation (FP) and backward propagation (BP), each worker acquires a batch of training data, propagates it through the model to get the loss function, and calculates the gradients of parameters based on the loss value; (2) gradients aggregation, all workers' gradients are aggregated (summation), and sent back to workers; (3) parameter update, each worker uses the aggregated gradients to update their parameters.

Two kinds of communication patterns are widely adopted for gradient aggregation.
 \begin{itemize}
    \item Parameter Server (PS)~\cite{ps}: This communication pattern requires specific parameter servers.  Each worker first sends its calculated gradients to parameter servers in each iteration, which then average the gradients and update the parameters, and finally broadcast the updated parameters to all workers.
    \item Ring All-Reduce~\cite{allreduce}: This communication pattern is a collective operation with two stages. A logical ring is formed among all workers. In the first stage, each worker sends a chunk of gradients to its right neighbor and receives gradients from its left neighbor, then accumulates its own copy to the received value, until all gradients are aggregated. In the second stage, each worker broadcasts its own aggregated chunk of gradients to other workers along the ring.
 \end{itemize}

Despite of scaling out the computation capacity, distributed model training suffers from the communication overhead~\cite{pipedream}. Every worker sends and receives tens MBs or even several GBs size of data in each iteration, which often becomes the bottleneck of the whole model training~\cite{poseidon,pipedream}. For example, PipeDream~\cite{pipedream} shows that communication overhead dominates the overall training time by over 90\% on 32 GPUs, and SwitchML reports a 29\% (for ResNet50) to 97\% (for DeepLight) communication time consumption in a 10Gbps network.


\parab{In-Network Aggregation (INA).} The advent of programmable switches provides an opportunity to scale out the distributed training~\cite{chen2018programmable}. The programmable switches typically use Reconfigurable Match-action Table (RMT) architecture with multi-stage pipelines. Each stage has memory (i.e., registers) to store durable network states and can load user-specified programs to process packets. Currently, a bunch of works apply programmable switches to accelerate distributed applications, e.g., NetCache~\cite{netcache}, NetLock~\cite{netlock}, and NetSeer\cite{netseer}.

\begin{figure}[htb]
    \centering
    \includegraphics[width=0.8\linewidth]{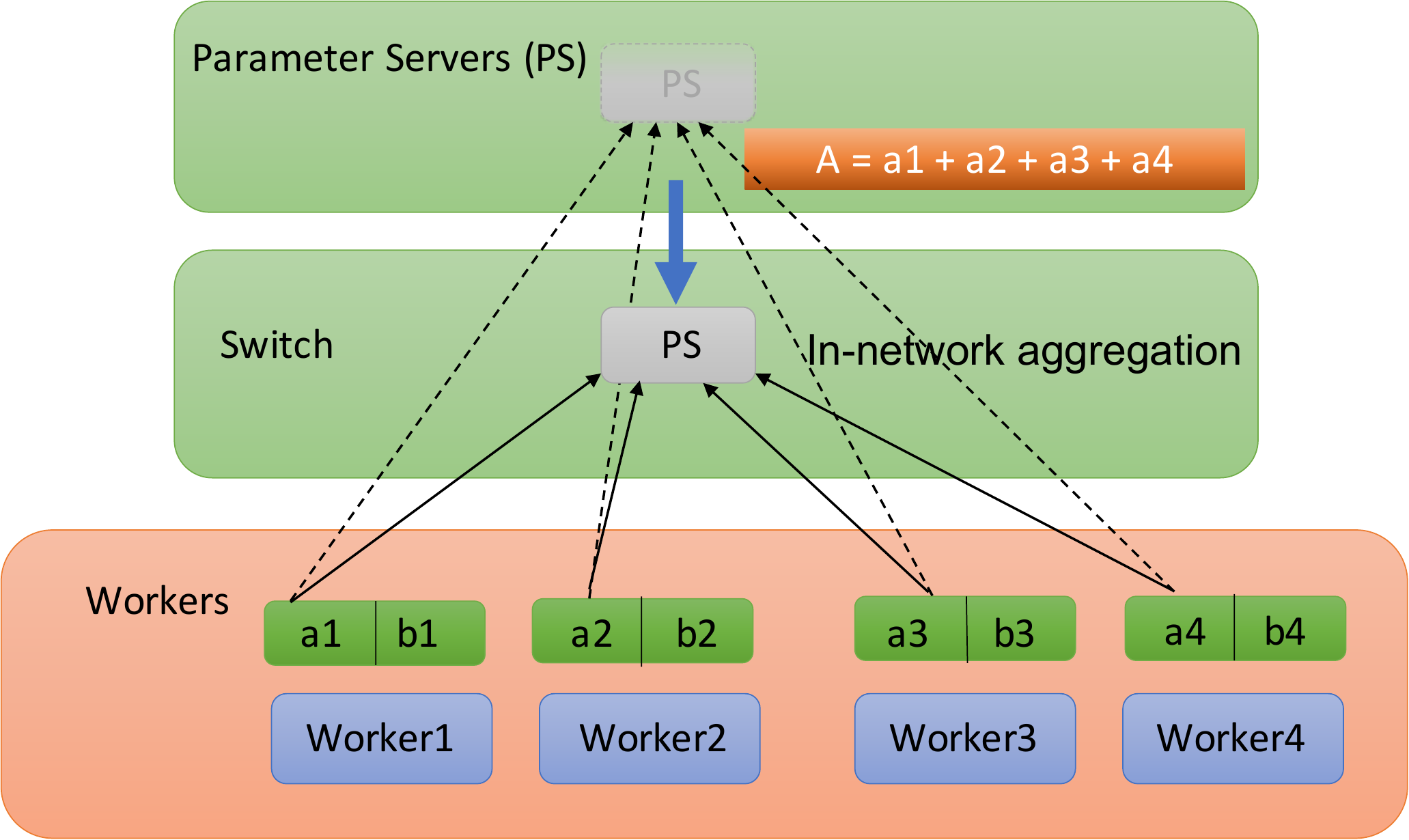}
    \vspace{-0.05in}
    \caption{An example of in-network aggregation.}
    \vspace{-0.1in}
    \label{fig:ina}
\end{figure}

Based on programmable switches, a recent technique called \emph{In-Network Aggregation (INA)} is proposed to scale out distributed training communication. The basic mechanism is illustrated in Figure~\ref{fig:ina}, where workers offload their gradient aggregation from hosts to the switch. As workers' gradients stream through the switch, the switch adds up the gradient packets from different workers as result packets, and streams the result packets either back to workers (e.g., SwitchML~\cite{switchml}, PANAMA~\cite{panama}) or to the PS (e.g., ATP~\cite{atp}). INA brings several advantages in the gradient aggregation (and the whole training process). First, INA reduces the traffic volume by letting the network carry the aggregation result instead of the raw gradients, and consequently eliminates the network bottleneck and reduces the communication time. Second, the aggregation operation is offloaded to the high-speed switches instead of servers, which not only saves CPU resources but also accelerates the aggregation speed. Third, the round-trip of a gradient packet and its result packet can be sub-RTT (except ATP with PS). Experiment results also show that INA can achieve up to $1.8\times$ speedup, compared to the current best practice, Ring All-reduce~\cite{switchml}.

\parab{Switch Memory Usage in INA (a.k.a. aggregator membership).} In INA solutions, the switch memory is organized as a pool of \emph{aggregators}, each is a basic unit to accumulate and store gradient packets. Gradient tensors in workers are organized as a sequence of packets, each with a unique sequence number. Gradient packets from different workers but with the same sequence number are assigned to the same aggregators and got aggregated, and the aggregation result is sent downstream; the whole process is called \emph{an aggregator task}. 

In the scenario of multiple jobs, existing solutions use switch memory in two ways. The first class statically partition switch memory and allocate a partition to a job (e.g., SwitchML~\cite{switchml}, iSwitch~\cite{iswitch}, PANAMA~\cite{panama}). The switch memory is not released until the job ends. The second class dynamically allocates switch memory to jobs (e.g., ATP~\cite{atp}). Each of the all aggregation tasks (from multiple jobs) choose its own aggregators in a decentralized manner (e.g., {\tt hash(jobID, seqNum)} in ATP), and would release the aggregator when the result packet (ACK) arrives at the switch; when two tasks collide in choosing an aggregator, the aggregator serve the first arrived one, i.e., First-Come-First-Serve (FCFS), and the later-arrived one falls back to the PS for aggregation.


\vspace{-0.1in}
\subsection{Switch Memory Bottleneck in INA}
\label{sec:motiv:ineffi}
\vspace{-0.1in}

\parab{Switch memory could not sufficiently support multi-jobs.} Switch memory is a scarce resource. The on-chip SRAM has to be designed of a small size to maintain high packet processing speed, e.g., ~10MB for Tofino~\cite{netseer}. In addition, production networks spare a portion of the switch memory for core switch functions such as forwarding table~\cite{switchml}, load balancing~\cite{miao2017silkroad}, firewall~\cite{cao2018cofilter}, etc. However, typical ML models have a size of hundreds of megabytes, which far exceeds the limitation of the switch memory.

Present programmable switch memory cannot support the multi-job setting in production networks. Theoretically, to support aggregating gradient packets at a line rate, the switch memory needed for a job equals the bandwidth delay product~\cite{switchml}, e.g., each job needs 1MB switch memory under 100Gbps bandwidth, therefore switchML can support at most ten jobs, which is insufficient for an industrial production environment, e.g., the trace of a two-month long workload from a GPU cluster in Microsoft~\cite{jeon2019analysis} shows that there can be a total of 96260 jobs, i.e., about a thousand jobs every day, and half of them last more than hours.
\parab{Non-preemptive aggregator allocation causes switch memory underutilized.} For the static INA solutions~\cite{switchml,panama,iswitch,hire}, the switch memory is not released until the job ends. Given training algorithms' nature of iterative computation and communication, the switch memory is idle without usage when the job is computing. 

Dynamic INA solutions (e.g., ATP) allocate aggregators to each aggregation task, but still disallow one job to preempt another one.
An aggregator's occupation time is from the arrival of the task's first packet to the leaving of its ACK. In aggregator occupation time, other aggregation tasks cannot use the aggregator.

The finest granularity for aggregator allocation is per aggregation, and such a granularity is still not sufficient to efficiently use the switch memory. The aggregator occupation time includes the gradient packet synchronization time and the round-trip time between the switch and the PS.

The synchronization could be postponed by straggler workers. There could be three reasons. First, different workers may not take exactly the same time to compute the gradients, especially under a heterogeneous environment. Second, system calls, memory copy and other overheads also contribute to the fluctuation of the start time of gradient sending. Third, the bandwidth obtained by different workers may also vary, especially in a multi-tenant cloud environment. 

The switch-PS round-trip time is introduced by the fallback mechanism. In INA solutions, the PS is used as the fallback to handle switch failure cases (e.g., packet loss). When the failure cases are of low probability and do not happen in the normal case, normally aggregated packets do not necessarily go through the round-trip, which contributes to the extra aggregator occupation time.




\parab{FCFS strategy ignores the differential switch memory demand of different jobs.}
Previous INA solutions~\cite{atp,switchml,panama} assign aggregators based on the arrival order of gradient fragment packets, which assumes all workloads have the same importance.
Therefore, there are more opportunities for performance gain if more ``important'' workloads can preempt less important ones. First, within a model, the front layer should preempt the other layers. In the training algorithm, the communication overlaps with the computation; and the forward propagation in one iteration can start immediately upon the arrival of the front-layer gradients in the previous iteration. Second, among jobs with diverse models, the models with higher communication overhead should have a higher priority. The reason is that when the computation time domains, the communication optimization only provides marginal benefit. Furthermore, when the overlapping is applied, reducing the communication time of a computation dominated model may not bring significant speedup. Third, for job scheduling, prioritizing jobs with the shortest remaining time can reduce the average job completion time (JCT).

\vspace{-0.1in}
\section{Key Ideas \& Challenges}
\label{sec:challenges}
\vspace{-0.1in}

To improve the memory utilization and average JCT, we propose two key ideas to tackle the limitations of existing INAs.

\parab{Data-plane Preemptive Allocation.} To make more efficient use of memory, we adopt a dynamic aggregator allocation scheme like ATP. When a later-coming gradient fragment packet is hashed to an already occupied aggregator, we allow preemption. Note that the above process occurs in the data-plane, so we need to design a data-plane preemption.

\parab{Co-designed Priority Scheduling.} Based on the above preemption mechanism, we tag priority for each gradient fragment packet. To enforce the priority scheduling for aggregation allocation (data-plane), we allow preemption if and only if the later-coming gradient fragment packets have higher priority. The end-host performs the priority calculation and tagging.

\vspace{-0.1in}
\begin{figure}[htb]
    \centering
    \includegraphics[width=0.95\linewidth]{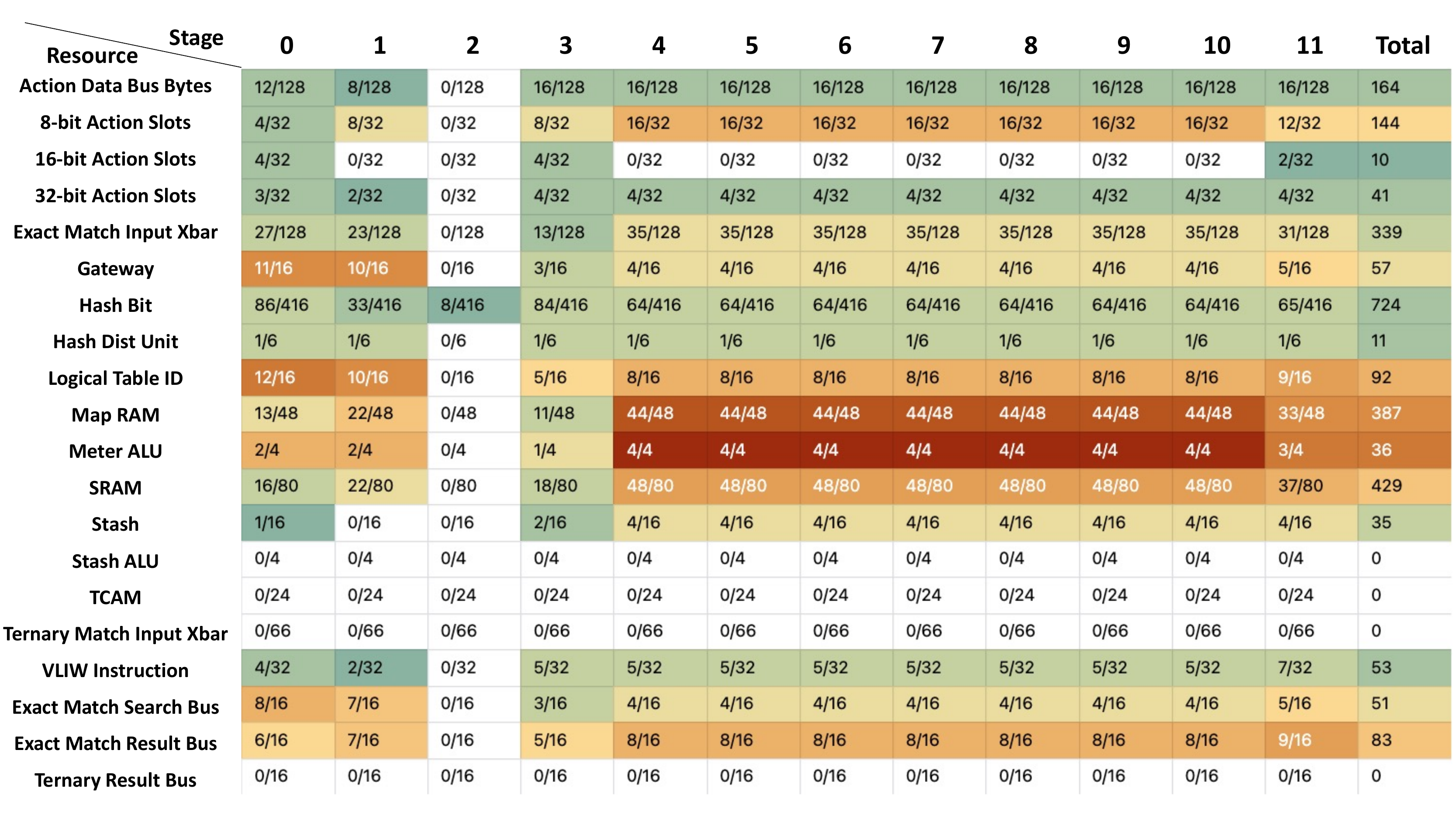}
    \vspace{-0.1in}
    \caption{Resource Summary of ATP P4 code~\cite{atp-code}, generated by P4i, a visualization tool for P4 programming~\cite{hauser2021survey}. This figure shows the occupation of 20 types of resources for every 12 stages. The shade of color reflects the resource usage. From stage 4 to stage 10, ATP uses up all Meter-ALU and more than $90\%$ map RAM.  }
    \vspace{-0.1in}
    \label{fig:challange:atp-resource}
\end{figure}
\vspace{-0.1in}

Although the above two key ideas look straight-forward, to implement the preemption allocation and priority scheduling on the programmable switch, we need to overcome three challenges.
First, how to implement the preemption mechanism purely in programmable switch pipelines without compromising performance? Programmable switches are limited in both computation and memory resources. Although existing INA solutions (SwitchML~\cite{switchml} and ATP~\cite{atp})  apply many optimizations, e.g., resubmit and recirculate, they still occupy most of the switch resources. As shown in Figure~\ref{fig:challange:atp-resource}, there is little space to add extra logic for preemption and priority strategy. Meanwhile, we cannot simply recirculate packets for more times, which sacrifices the throughput.

Second, to avoid impacting the training accuracy, the protocol needs to guarantee \emph{all-case} correctness of the aggregation results. Since switches in INA are stateful, and various network conditions, e.g., packet loss, could happen in the whole duration, as a result, the protocol needs to handle more complicated abnormal transmission cases (e.g., partial aggregation and memory leak~\cite{atp}). Previous INAs~\cite{atp,switchml} have tackled the reliability and correctness issues by using bitmaps to record the already arrived packets. However, when enabling preemption on switch aggregators, existing reliability will not work. Preemption will kick out the previous gradients together with the bitmap. (You can keep the old bitmap in the aggregator, however, it will cost more memory and logic resources.) Meanwhile, when enabling preemption, the reliability mechanism needs to cover more cases, e.g., the transmission of previous partial aggregation result.

Third, the protocol needs to decide a preemption frequency. When a preemption occurs, the switch will send the partial aggregation result in the fallback PS. Therefore, too frequent preemptions will increase in-network traffic volume and impact the performance of INA. However, if preemption is too conservative, aggregators with partial results may spend too much time on waiting for straggler packets, wasting time in idle and still degrading the performance. 

\vspace{-0.15in}
\section{Overview of \sys}
\vspace{-0.1in}
\parab{Goal.} 
In this paper, we aim to improve switch memory utilization and enforce priority allocation strategy so as to promote multi-job training performance. Our intuition is to introduce a preemption mechanism into the aggregator allocation so as to reduce aggregator idle time. We design a system named \sysname for our goal.



\parab{Architecture Overview.} 
\sysname consists of a dedicated transport layer protocol to support INA with preemption enforcement, and a preemption policy for jobs. In the protocol, the switch performs the major aggregation functionality, and a host-based PS works as the fallback. \sysname dynamically allocates aggregators to aggregation tasks. Each aggregation task has a priority. The aggregator supports priority-based preemption: when two tasks collide at an aggregator, if the latter one has a lower priority, it would pass through to the PS; otherwise, it evicts the early task from the aggregator to the PS, and seizure the aggregator. The fallback PS only handles exceptional cases, e.g., packet loss and preemption.

According to \sysname's priority policy ($\S$\ref{sec:design:priority}), an aggregation task's priority is computed by combining its layer position in the neural network, its communication-to-computation ratio, and its remaining execution time (or attained service time if the remaining time is not available). The priority is encoded into each gradient packet header for the switch aggregator to make the decision about preemption.

\begin{figure}[!ht]
    \centering
    \includegraphics[width=0.99\linewidth]{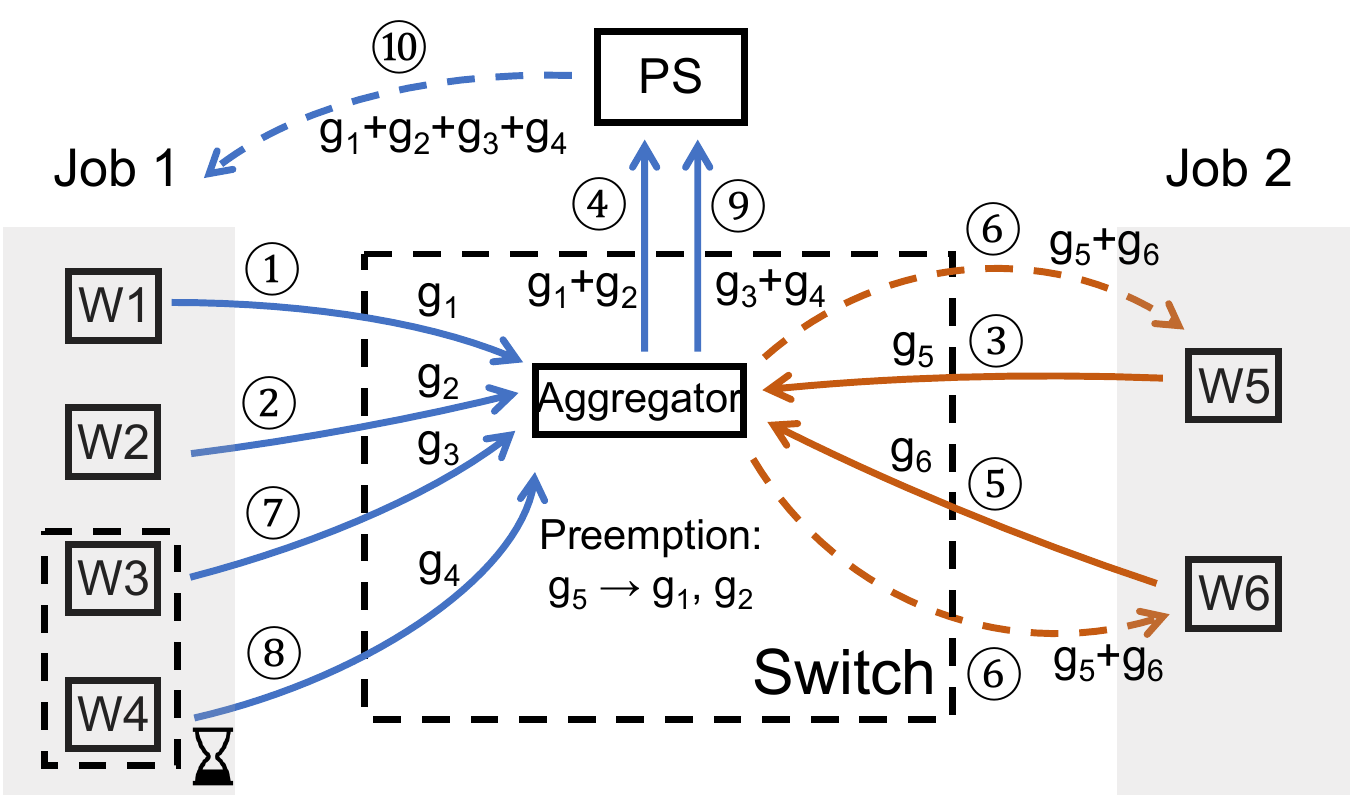}
    \vspace{-0.05in}
    \caption{An example of preemptive aggregator allocation scheme. The blue (red) arrows represent the flows of Job 1 (2). The dotted arrows represent the multicast flows. We note the gradients of Worker $i$ as $g_i$. Worker 3 and 4 are the stragglers. The serial numbers show the sequence of each flow event. }
    \vspace{-0.1in}
    \label{fig:meth:preempt}
\end{figure}

\parab{Example.} Figure~\ref{fig:meth:preempt} exemplifies the preemptive allocation procedure.
Job 1 has four workers, W1, W2, W3 and W4. Suppose W3 and W4 are the stragglers in this iteration. Firstly, W1 and W2 send the gradient packets to switch (\textcircled{\scriptsize 1}\textcircled{\scriptsize 2}), since the corresponding aggregator is empty, the aggregator is allocated to Job 1 and waits for the arrival of the gradient packets of W3 and W4.
At this time, Job 2 with two workers, W5 and W6 start to send gradients (suppose $g_5$, $g_6$ have higher priority than $g_1$, $g_2$, $g_3$, $g_4$). Upon the arrival of gradient packet of W5 (\textcircled{\scriptsize 3}), the switch performs preemption allocation: the switch sends the partial aggregation result of W1 and W2, i.e., $g_1+g_2$, to the PS and releases the aggregator (\textcircled{\scriptsize 4}). After the arrival of $g_6$ (\textcircled{\scriptsize 5}), the aggregation of Job 2 is completed, so the switch multicasts the result $g_5+g_6$ to W5 and W6, and releases the aggregator (\textcircled{\scriptsize 6}). When $g_3$ and $g_4$ from W3 and W4 arrive (\textcircled{\scriptsize 7}\textcircled{\scriptsize 8}), the aggregator is allocated to Job 1 again, the aggregation result $g_3+g_4$ will be send to the PS (\textcircled{\scriptsize 9}).\footnote{We use a reminder mechanism to enforce this deallocation, see $\S$\ref{sec:design:endhost}.} Finally, the PS adds up the two partial aggregation results of Job 1 ($g_1+g_2+g_3+g_4$) and PS multicasts them to W1,W2,W3 and W4 (\textcircled{\scriptsize 10}).




\parab{Discussion.} Although the preemptive allocation will introduce additional aggregation packets for some gradients, from the perspective of all DLT jobs, de facto preemption is able to reduce the in network traffic volume.
Each aggregation computation (i.e., gradient summation) in switch reduces one packet in the network. Therefore, to reduce the traffic volume, we should increase the number of aggregation computations per unit time. Recall that under non-preemptive allocation, the aggregator will be idle when it is waiting for the late coming workers' gradients. During this period, other gradients mapped to the aggregator will back off to the normal PS mode, which brings lots of in-network traffic. 
With preemption, we can reduce the times of backing off to the normal PS by introducing only several partial aggregation packets. 


\vspace{-0.1in}
\section{System Design}
\vspace{-0.1in}

\sys codesigns the end-host and the switch. 
On the worker side of the end-host, like ATP, \sys divides each gradient tensor into gradient fragment packets with the same size and marks the location information (i.e., sequence number) on the header. \sys also ensures that for the workers in the same job, each gradient fragment packet at the same location has the same sequence number.

On the switch side, all \sys jobs share the switch memory (aggregators). Therefore, jobs may content for aggregators, unlike ATP, \sys allows preemption. Considering the limited switch computation resources, \sys applies a technique we call \textit{packet swapping} under preemption: when a new gradient packet preempts an aggregator, we swap the aggregated gradient value in the aggregator with the payload of the packet, then send the packet to the corresponding PS. 

On the PS side of the end-host, \sys performs further aggregation for the gradient fragment when: (a) The gradient fragment was preempted during the aggregation in switches. (b) The gradient fragment has a hash collision during the aggregator allocation and fails to preempt with a lower priority. (c) Packet loss occurs during the aggregation.

\vspace{-0.1in}
\subsection{End-host Logic}
\label{sec:design:endhost}
\vspace{-0.05in}

Workers in \sys tag priority at the header of gradient fragment packets and push gradients to the switch. If there is no preemption nor packet loss, workers will pull parameters from the switch. Otherwise, they pull parameters from the PS, which is used to guarantee the correctness of the gradient aggregation for the corner cases.

\parab{Packet Format.}
The \sys Header adds an 8-bit priority field based on the ATP Head. Recall that a ATP Header also contains: 1) two bitmaps, bitmap0 to record the place of the first-level switch and bitmap1 to record the second-level switch. 2) job ID and sequence number. 3) aggregator index. 4) gradient fragment. We will use these fields later.
\sys has a special packet called reminder packet, it is a special type of \sys gradient fragment packet, all fields,   except the job ID and sequence number, are 0.

\parab{Worker Pushing Gradients.}
\sys provides an interface to the upper DLT applications for pushing gradients. DLT jobs need to provide the gradient tensor to be pushed, the layer of the tensor, the job's remaining time (or the time already in progress) and the ratio of computation to communication measured from the last iteration. Note that these information are readily accessible.
Upon the interface is called, \sys will first calculate the priority according to the formula in $\S$\ref{sec:design:priority}. Since the priority field has only 8 bits, we need to compress the priority into a 8-bit fixed-point, we omit the compression detail here, since it is similar to the float-point gradients converting to fixed-point. In addition, we calculate the hash value of the current gradient fragment according to the job ID and sequence number, the hash function is the same as ATP's. The hash value is tagged on the aggregator index field of the \sys header. Since most of the programmable switches do not support floating point calculation, like SwitchML and ATP, \sys will convert floating point gradients to fixed point at the end-host.

\parab{PS Assisting with Aggregation.}
Without packet loss nor preemption, INA can do the aggregation completely in switches. However, in the real case, we need to take these corner cases into account to ensure correctness. Due to the limited resources of switches, we move the functions of guaranteeing aggregation correctness to the PS.
For each job, PS allocates a memory space to cache the intermediate results of the aggregated gradient fragment. The space is in the form of a dictionary, and the key is sequence number and the value is {\tt<bitmap, aggregation result, timestamp>} (note that the bitmap contain aggregation completion information of all workers). PS will assist in aggregation in the following three cases.

\vspace{-0.1in}
\begin{figure}[!ht]
    \centering
    \includegraphics[width=0.9\linewidth]{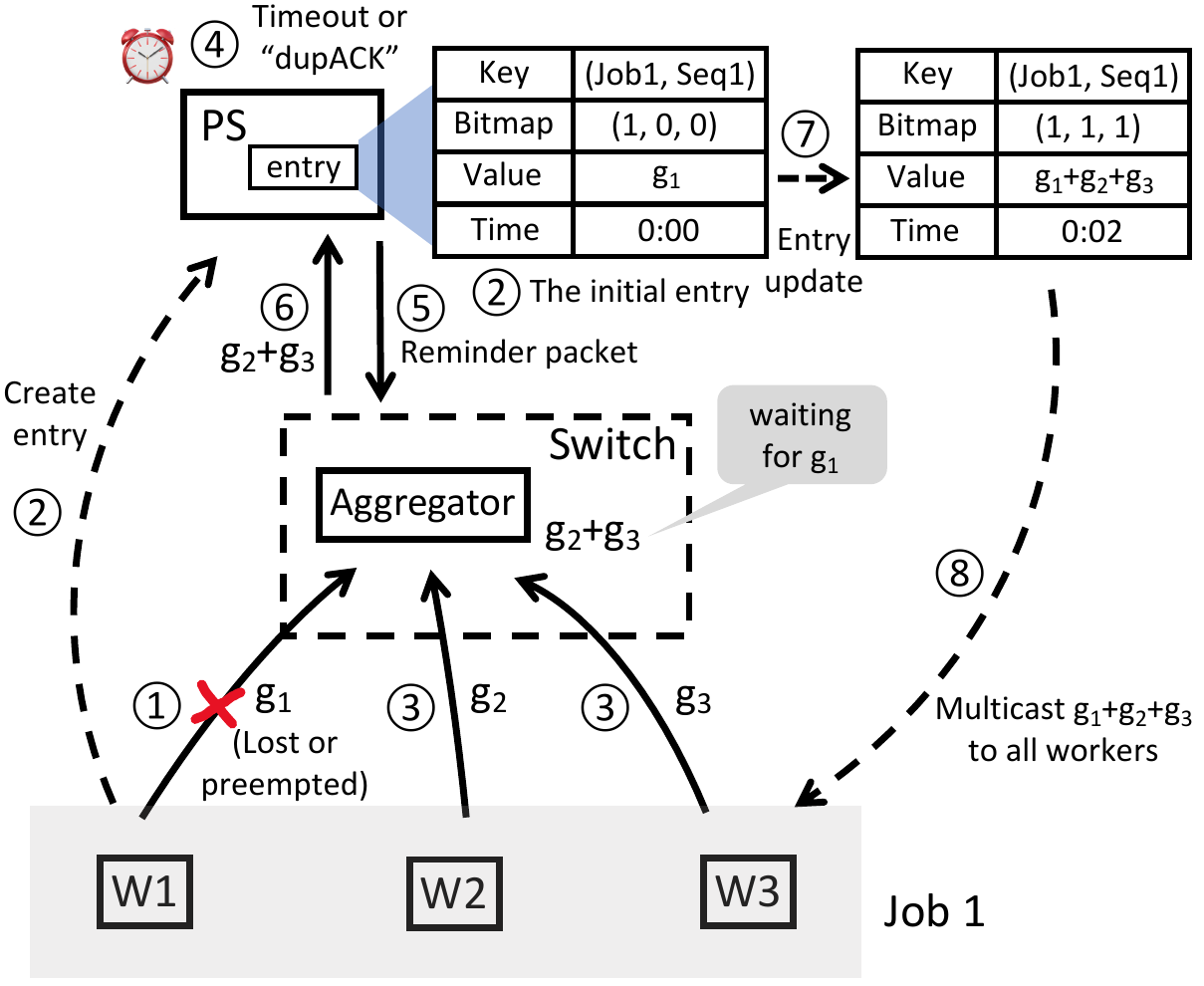}
    \vspace{-0.1in}
    \caption{An example of reminder mechanism. The solid arrows represent the flows in the network. The dotted arrows represent other events. Job 1 has three workers. We note the first gradient fragment of Worker $i$ as $g_i$. Here is the workflow: $g_1$ was sent out first, however, due to reasons like packet loss, preemption, failing to preempt the aggregator, $g_1$ lost possession of the aggregator (\textcircled{\scriptsize 1}). Upon the worker or the PS detect the failure (details in $\S$\ref{sec:design:loss}), the PS will create an entry for this gradient fragment (\textcircled{\scriptsize 2}). When $g_2$ and $g_3$ were sent out and reached the switch (\textcircled{\scriptsize 3}), suppose they occupied the aggregator, then they will wait for $g_1$. However, $g_1$ had been sent, so the switch cannot do deallocation by itself. Therefore, \sys applies a reminder mechanism to assist the deallocation. When the PS detects a timeout for an entry (2s in this example, practical setting in $\S$\ref{sec:impl}), or receive three consecutive aggregated gradients other than the expected sequence number, we called it as "dupACK" (\textcircled{\scriptsize 4}), it will send a reminder packet to the switch, which contains the job ID and sequence number (\textcircled{\scriptsize 5}). The reminder packet will fetch the aggregated result (packet swapping, see $\S$\ref{sec:impl}), i.e, $g_2+g_3$, from the switch (\textcircled{\scriptsize 6}). Then the PS will update the corresponding entry (\textcircled{\scriptsize 7}). Finally, if the bitmap of the entry are all "1", the PS will multicast this aggregated result to all workers  (\textcircled{\scriptsize 8}). }
    \vspace{-0.1in}
    \label{fig:meth:reminder}
\end{figure}
\vspace{-0.1in}


\textit{The gradient fragment has been preempted.} When a preemption occurs in an aggregator, the original gradient fragment is kicked out and the previous value and bitmap are sent to PS in order not to waste the previous computation. After receiving the partial aggregation result of a gradient fragment, if the PS dictionary of the corresponding job does not have the entry mapped by this sequence number, the PS will create a key-value pair; if it does, the PS will add its value to the corresponding entry. Since the gradient fragment has already sent some aggregation results to the PS, next time it comes to the switches, it will wait for the previously completed gradient fragment and cannot complete the aggregation on the switch. To solve this issue, for a gradient fragment in the dictionary, the PS will send \textit{reminder packets} to the corresponding switches when it has not received a gradient fragment of the same job and with the same sequence number for more than a timeout , or when the PS receives three gradient fragments with larger sequence numbers ("dupACK"). Figure~\ref{fig:meth:reminder} illustrates the reminder mechanism\footnote{Actually the reminder mechanism is just the last line of defense. A long time occupied aggregator will be more likely kicked out by the later coming gradients, before the reminder mechanism makes effect.}.

\textit{The gradient fragment has tried to preempt but failed.} This happens when the later coming gradient fragment has a lower priority than the existing one in the aggregator. For such gradient fragment, the switch will send it directly to the correspond PS. When the PS receives such gradient fragment, it will treat the gradient fragment as a partial aggregation result, then the following process is as same as case 1.

\textit{Packet Loss.} When there is a packet loss, unlike ATP, all retransmitted packets will be directly sent to the PS via reliable transmission, e.g., TCP. We do not chose to send to the switch due to the following two reasons: 1) When enabling preemption, the switch may not have the whole bitmap of a gradient fragment. Therefore, the switch cannot judge whether the gradient fragment has already been aggregated. 2) Packet loss is rare in the data center, and we apply a selective retransmission mechanism, only the workers who lost packets are required to resend the gradients. Therefore, directly sending to the PS will not introduce much traffic volume. We will cover this part in detail in $\S$\ref{sec:design:loss}. 

\parab{Worker Pulling Parameters.}
\sys workers receive parameters from switches or PSes. \sys applies a window-based packet sending method, that is, after the initial window of packets is sent out, workers are waiting for the parameter packets. When receiving a parameter packet, the worker checks whether it has the expected sequence number, that is, the first sequence number in the sending window. If it is, the worker will send the next gradient fragment packet. We use the same initial window size (60KB at 100Gbps) and congestion control algorithm applied ATP~\cite{atp}. As PSes do not handle all gradient fragments, in some cases of packet loss, workers also need to tackle the packet loss and retransmission, which we will introduce in $\S$\ref{sec:design:loss}.

\vspace{-0.1in}
\subsection{Switch Logic}
\label{sec:design:switch}
\vspace{-0.1in}

\begin{figure}[!ht]
    \centering
    \includegraphics[width=0.92\linewidth]{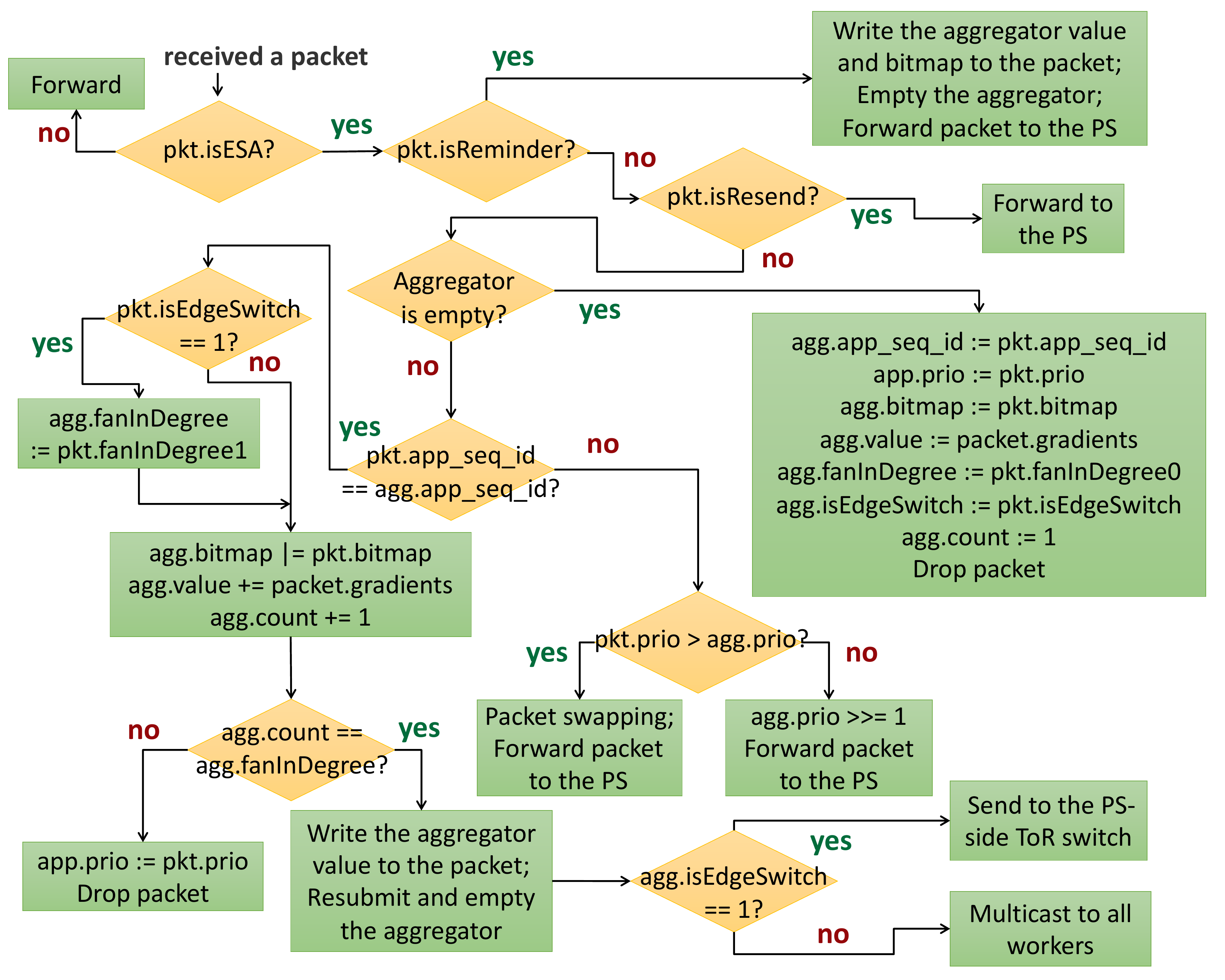}
    \vspace{-0.1in}
    \caption{Pseudocode of the switch logic. Here, agg is short for aggregator and pkt is short for packet.}
    \vspace{-0.15in}
    \label{fig:design:switch}
\end{figure}


\parab{Switch Memory Layout.}
The switch layout of \sys is based on ATP, and we add an 8-bit priority field to each aggregator. Each aggregator also contains 1) 32-bit bitmap, 2) 32-bit counter, 3) Job ID and sequence number, 4) Fan-in degrees for the first-level and second-level switch, 5)1-bit boolean indicating the aggregation level, 6) aggregator value. These fields will be used later in the design description.

\parab{Preemptive Aggregator Allocation.}
Recalled that ATP applies a hierarchical aggregation with the first-level switch at the workers' rack and the second edge switch at the PS's rack. \sys also adopts this hierarchical aggregation to support multi-rack aggregation.

As shown in Figure~\ref{fig:design:switch}, 
when a gradient fragment packet arrives at the switch, the switch will first check whether the corresponding aggregator is empty or not, if it is empty, then the switch will directly assign the aggregator to the gradient fragment, i.e., write the job ID and sequence number of the aggregator and initialize the aggregator value, bitmap, and counter, and record the aggregation level. The fan-in degree is initialized as the worker number of first-level aggregation. 
If the corresponding aggregator is not empty, the switch will  compare the job ID and sequence number of the aggregator with the gradient fragment packet. If both are the same, the switch will first check the aggregation level and then aggregate the gradient fragment and update the bitmap, and counter, meanwhile the switch will check whether the gradient packets of all workers have arrived, and perform priority renewal or multicast. If not, the switch compares the priority of the aggregator with the gradient fragment packet's. If the priority in the aggregator is higher or equal, the preemption will fail and the current gradient fragment packet will be sent to the PS. Meanwhile, the aggregator performs \textit{priority downgrading}, i.e., priority value divided by 2 (shift one bit to the right). Otherwise, the preemption operation is performed, and the new gradient fragment will obtain this aggregator, the current aggregation result of the old gradient fragment will be sent to the PS. We use a technique we call \textit{packet swapping} to simplify the preemption logic, which we will introduce in $\S$\ref{sec:impl}.

\parab{Aggregator Deallocation.}
The aggregator deallocation will be executed under the following three scenarios : 1) The gradient fragment in the aggregator is preempted; 2) All worker's gradient fragment packets have arrived; 3) Switch receives a reminder packet from the PS.
In case 1, 3, the current aggregation result will be sent to the PS, then the switch deallocates the aggregator. In case 2, for the first-level switch, it will send the aggregation result to the second edge switch. For the second edge switch, it will multicast the result to all workers directly.



\vspace{-0.1in}
\subsection{Dealing with Packet Loss}
\label{sec:design:loss}
\vspace{-0.05in}

\sys should consider the loss of gradient fragments or parameter packets since it rebuilds the transport.
Although the reliability mechanism in ATP~\cite{atp} and SwitchML~\cite{switchml} have already considered the packet loss recovery, we cannot directly adopt their loss recovery mechanisms to \sys. The reasons are as follows:

\begin{itemize}
	\item To make sure no retransmitted gradient packet be repetitively aggregated, SwitchML and ATP use bitmaps to record every successful aggregated gradient packet. When enabling preemption, due to the logic and memory limitation, switches cannot store the bitmap of preempted gradient fragments. Without the  bitmap, switches cannot judge whether a gradient packet is a resent one or not.
	\item Preemption incurs more possible scenarios for aggregator allocation, e.g, success or failure of preemption. Therefore, \sys needs to design the reliability mechanism for the new emerging cases.
\end{itemize}


Here, we will discuss how \sys deal with the packet loss. We have made the following classification based on following three conditions: Is there a hash collision? Is there a  preemption? Where the packet loss occurs? In the following statement, for the end-host to end-host transmission, we use reliable transport protocols, e.g, TCP.

\parab{Case 1:} \textit{No hash collision, and the packet loss occurs when a gradient fragment packet is sent to the switch.} Since there is no hash collision, the PS has no information about the gradient fragment. Therefore, for this gradient fragment, no reminder packet will be sent to the switch from the PS. To solve the issue, we apply a similar timeout mechanism in the workers as the PS. When the worker receives three gradient fragments with larger sequence number than the expected, or when timeout occurs, the worker will send a reminder packet to the PS, then PS will create an entry for the current gradient fragment. Next, the PS will immediately send a reminder packet to all related switches. 

\parab{Case 2:} \textit{No hash collision, and the packet loss occurs when it is multicasted back to workers from the switch.} There are two possible scenarios for this case. 
First, some of the workers did not receive the parameter. To avoid the repetitive computation, we set up a cache in each worker to store the recently received gradient fragment, the cache size is the same as the window size. For the workers which didn't receive the parameter, they will send a reminder packet to the PS. Then PS will send a reminder packet to switch and query packets to all workers to check whether they have received the parameter. If the worker has received the parameter, it will return the value of the parameter to the PS, and PS sends the parameter to the not-yet received workers.   
Second, none of the worker received the parameter (the probability is very low). In order to ensure the aggregation of this gradient fragment can complete in the next time, we use PS to aggregate this gradient fragment instead.


\parab{Case 3:} \textit{A hash collision occurs and no preemption occurs, but the failed gradient is lost when it is sent to the PS.} In this case, the failed gradient fragment will not be able to create an entry on the PS, but the alert mechanism on our worker side can solve this problem.

\parab{Case 4:} \textit{With hash collision and preemption, and the preempted gradient fragment packet is lost when it is sent to the PS.} In this case, the preempted gradient fragment will not be able to create an entry on the PS, nevertheless, the reminder mechanism on the worker side can tackle this issue.

\parab{Case 5:} \textit{With hash collision and preemption, and packet loss occurs when the gradient fragment packets of other workers with the same sequence number are sent to the switch.} If the gradient fragment has an entry on PS, the reminder mechanism on PS can be also handle this case. If not, worker side reminder mechanism will tackle the packet loss.

\vspace{-0.1in}
\subsection{Preemption Policy}
\label{sec:design:priority}
\vspace{-0.05in}


To minimize the average JCT, we enforce a priority-based aggregator allocation scheme. We observe three aspects of the priority relations among different gradients. First, from an \textit{intra-model} perspective, gradients from the front layers matter. We can obtain the layer information from the model. Second, from an \textit{inter-model} perspective, gradients from the models with higher communication overhead matter. We can measure the time of communication and computation from the previous training iteration. Third, from an \textit{inter-job} perspective, gradients from models with the shortest remaining time matter. If we have prior knowledge about the overall training time for a job, we can calculate the remaining time. However, in most case, the training time is agnostic, we will estimate it by using the service the job has attained so far. Meanwhile, to improve the aggregators' utilization, we need to degrade the priority of the gradients which has occupied the aggregators according to the occupation time. Here, we will describe how to calculate the priority of each gradient.


\begin{table}[htbp]
\begin{tabularx}{0.48\textwidth}{p{0.07\textwidth}X}
\toprule
  $J$ & DLT jobs \\
  $l$ & layer of the gradients \\
  $P_j(l)$ & the priority of the gradients in layer $l$ for Job $j$ \\
  $T_j$ & remaining time to convergence for Job $j$ \\
  $L_j$ & number of layers in the model of Job $j$ \\
  $Comm_j$ & communication overhead of Job $j$ \\
  $Comp_j$ & computation overhead of Job $j$ \\
\bottomrule
\end{tabularx}
\vspace{-0.05in}
\caption{Definitions and notations}
\vspace{-0.2in}
\label{sec:method:notation}
\end{table}


\parab{Priority setting: } We use the notations as the table~\ref{sec:method:notation} shows.
For each gradient, we set the priority according to the job, the model, and the layer in which it is located. The formula is:
\vspace{-0.05in}
\begin{equation}
P_j(l)=\frac{1}{T_j}\times\frac{L_j}{l}\times\frac{Comm_j}{Comp_j}
\end{equation}
\vspace{-0.15in}

Intuitively, the priority setting reflects the three observations we mentioned above. We simply multiply the three factors together due to two reasons. First, using the product form eliminates the need to consider the units of each variable and doing normalization. Therefore, each job can calculate its own priority independently at the end-host. Second, The above three factors come from different levels, thus we chose multiply instead of summation.

\parab{Priority Downgrading: } To avoid an aggregator be occupied for a long time, we enforce a priority downgrading mechanism at the switch. Considering the scarce computing resources in the switch, we only downgrade the priority of the gradients in the aggregators under the hash collision.
If there is a hash collision without preemption, the priority of the gradient in the aggregator is halved, i.e, shift one bit to the right.
    \vspace{-0.1in}
\section{Implementation}
\label{sec:impl}
\vspace{-0.1in}

In this section, we discuss two implementation details of \sys.

\parab{Packet Swapping.}
Two base operations are required for the aggregator upon preemption. First, the current aggregation result of the old gradient fragment should be fetched and stored. Second, the aggregator should be reallocated to the new gradient fragment. Note that we cannot manipulate a stateful memory unit, e.g., aggregators, twice in one pass. Fortunately, the register type of P4 can read and write the stateful memory in one operation. Leveraging this feature, we "swap" the aggregator registers with the gradient fragment of the later-coming packet. Suppose one gradient entry in the packet is {\tt esa\_gradients.data0}, we use the value to update the register {\tt (update\_lo\_1\_value : esa\_gradients.data0)}, and use the previous register value to update the gradient entry of the packet {\tt (output\_dst : esa\_gradients.data0; output\_value : register\_lo;)}. Finally, the current gradient packet with the previous gradient fragment's aggregation result is forwarded to the PS.
For the metadata fields like appID and priority, we apply the same swapping method. However, these fields are manipulated in the early stages to fetch the necessary information of the aggregator. Due to the limitation of the switch logic, we cannot manipulate them twice at the later stages, i.e, performing swapping. Instead, we perform resubmit, i.e., redirect the packet from the deparser of the ingress to the parser, to handle these cases.



\parab{Setting of Reminder Mechanism.} We have reminder mechanisms on both the worker and the PS. A reminder packet will be sent out upon timeout or three consecutive packets with larger sequence numbers than the expected one. Our calculation of timeout takes reference from the TCP timeout~\cite{kesselman2005optimizing}. For the worker, the "RTT" is the interval of sending the gradient fragment packet and receiving the parameter packet. For the PS, the "RTT" is the interval of entry setup and aggregation completion. To avoid the spurious reminder, we also set a "RTO\_min", the value is 1ms in our setting. 

    \vspace{-0.2in}
\section{Evaluation}
\label{sec:evaluation}
\vspace{-0.1in}

We evaluate \sys with a combination of testbed validation and simulation measurement. The key findings are as follows:
\begin{itemize}
	\item We evaluate the end-to-end DNN training of \sys on our testbed. BytePS+\sys can achieve up to 1.27$\times$/1.15$\times$ training speed up compared to BytePS/BytePS+ATP.
    \item We measure the JCT of \sys under the 64-nodes simulation. \sys can improve the JCT by up to 1.89$\times$/1.35$\times$ compared to SwitchML/ATP.
    \item We deep dive the effectiveness of preemption on memory utilization and priority scheduling. \sys can significantly improve the utilization by up to 2.27$\times$ and further improve the JCT by up to 1.16$\times$ with the scheduling.
\end{itemize}

\vspace{-0.1in}
\subsection{Testbed Experiments}
\label{sec:eval:testbed}
\vspace{-0.05in}


We integrate \sys into BytePS~\cite{byteps}, a state-of-the-art DNN training framework and evaluate its end-to-end performance on a small-scale testbed. We also use microbenchmarks to evaluate the communication speedup.

\vspace{-0.1in}
\subsubsection{Testbed Setup}
 \vspace{-0.1in}
 
%
%
\parab{Testbed:} Our testbed contains 5 physical GPU servers, each with 2 V100 GPUs, 40 CPU cores (Intel Xeon Gold 5115), 128GB memory, 2 Mellanox ConnectX5 100Gbps NICs, and one Edgecore Wedge100BF-32X switch, each server have two 100Gbps links connecting to the switch. To scale up our testbed, we further divide one physical server into two separated docker containers, each with $1\times$ GPU, $20\times$ CPU cores, 64GB memory and a 100Gbps virtual NIC. 

\parab{Models and Workloads:} We chose the representative models used in ATP, i.e., ResNet50~\cite{resnet2016cvpr} and VGG16~\cite{vgg} training on the Cifar100~\cite{cifar}, a widely-used dataset. We also run a microbenchmark which only involves the communication. For the microbenchmark, each worker repeatedly transfers fixed size tensors to be aggregated and then sent back.

\parab{Baselines and Metrics:} To support the end-to-end DNN training, all INA algorithms are integrated into BytePS~\cite{byteps}. To exclude the impact of the end-host network stack, we apply TSO and MP-QP features to all algorithms. We compare \sys against ATP and vanilla BytePS (NtoN RDMA). We use time to epoch (TTE) and time to accuracy (TTA) for the end-to-end training.
The baselines of microbenchmark are ATP and SwitchML.
The metric is the same in ATP~\cite{atp}, i.e., aggregation throughput. Packet size of ATP and \sys is 306B and SwitchML is 180B. SwitchML jobs evenly share the memory.

\vspace{-0.1in}
\subsubsection{End-to-End DNN Training} 
\begin{figure}[htb!]
    \centering
    \subfigure[TTE]{
        \includegraphics[width=0.472\linewidth]{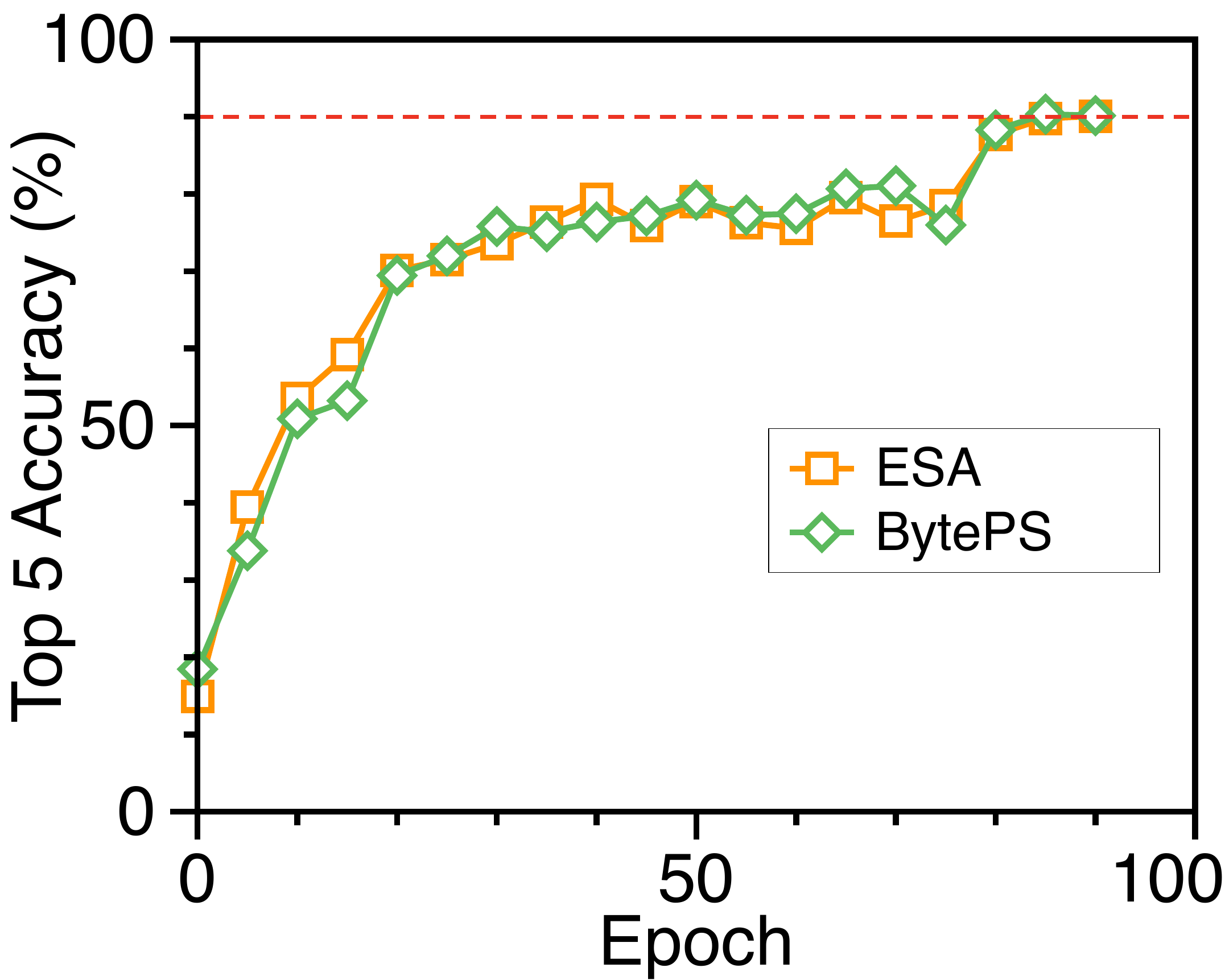}
    }
    \subfigure[TTA]{
	\includegraphics[width=0.472\linewidth]{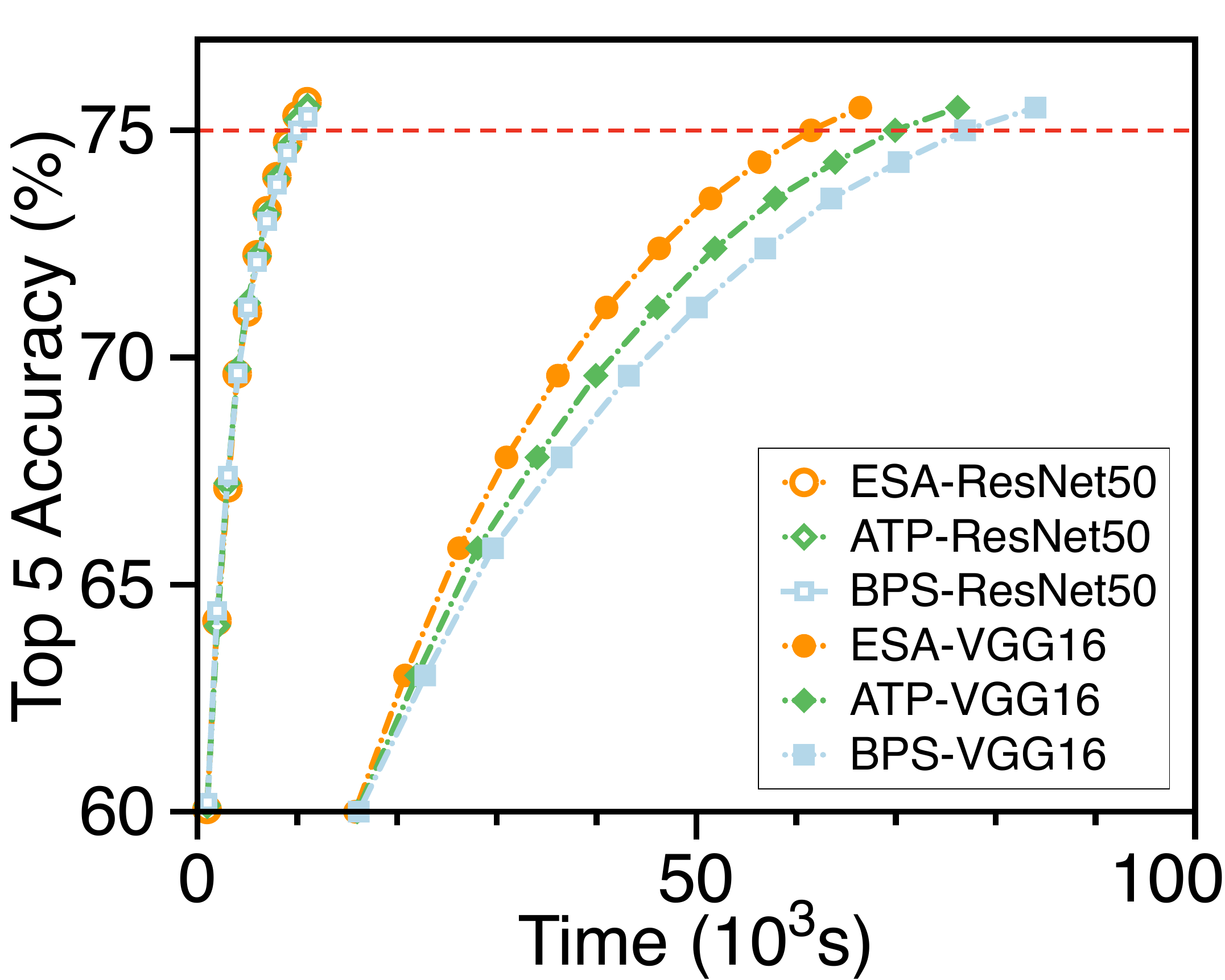}
    }
    \vspace{-0.1in}
    \caption{End-to-End DNN Training}
    \vspace{-0.1in}
    \label{fig:testbed:tta}
\end{figure}

We first verify that \sys does not affect the training accuracy by conducting a single job with 8 workers training with ResNet50. Due to the fact of our small scale testbed, we limit the switch memory (aggregators) for our INAs to 1MB. In Figure~\ref{fig:testbed:tta} (a), we can see that the curve of \sys is similar to BytePS's and they convergence to the small accuracy. We further evaluate \sys in a multi-tenant setting, we run two DNNs, ResNet50 and VGG16, each with 4 workers. The batch size is 32. For \sys and ATP, we set one PS for each job. As shown in Figure~\ref{fig:testbed:tta} (b), for VGG16 \sys reaches 75\% top-5 accuracy $1.15\times$ and $1.27\times$ faster than the ATP and BytePS. For ResNet50, \sys only slightly outperforms the previous solutions (< $1.01\times$), as ResNet50 is computation-intensive, which is consistent with the observations in ATP~\cite{atp}.

\vspace{-0.1in}
\begin{figure}[htb!]
    \centering
    \subfigure[Different Tensor Size]{
        \includegraphics[width=0.472\linewidth]{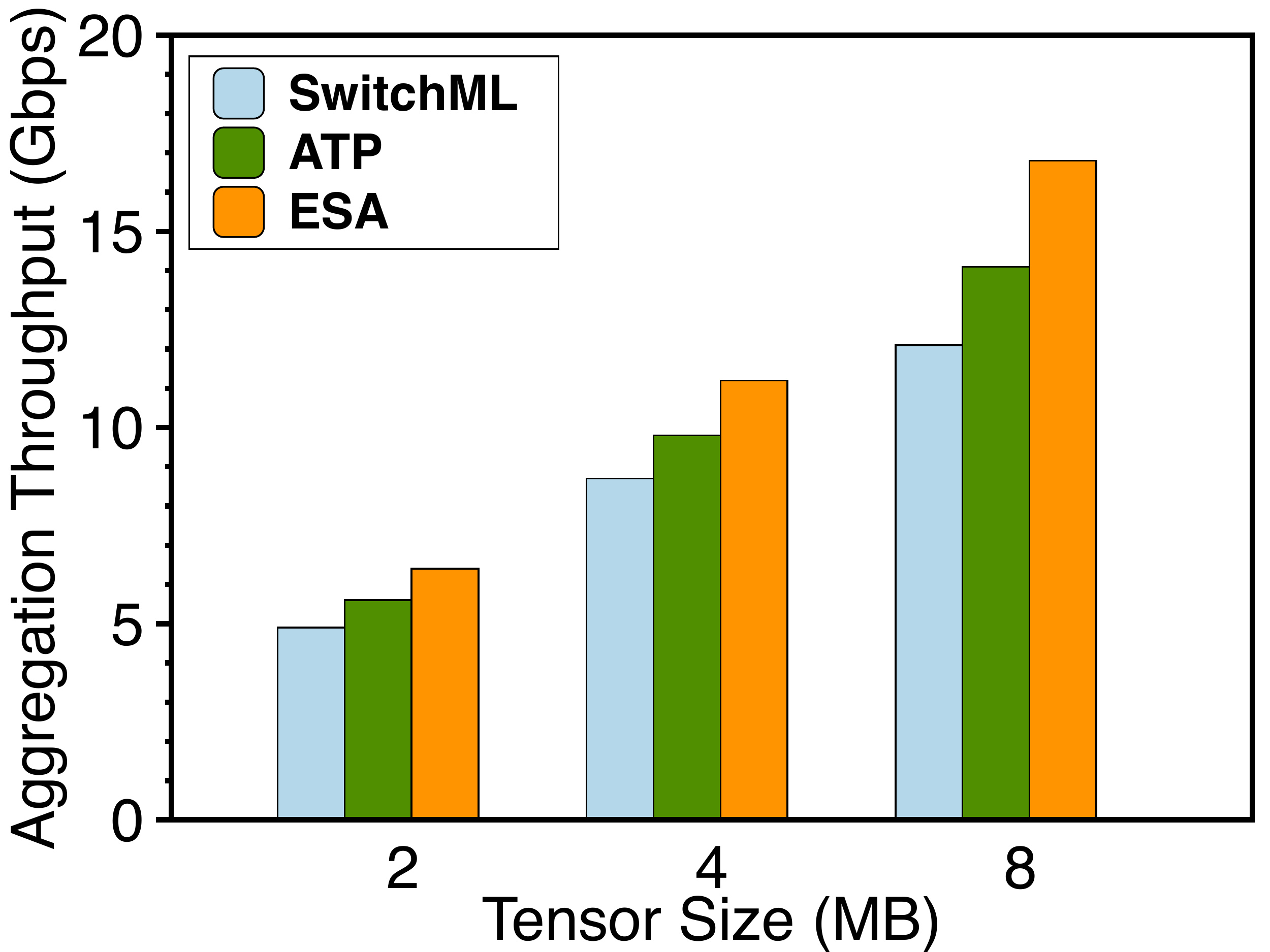}
    }
    \subfigure[Different Job Number]{
	\includegraphics[width=0.472\linewidth]{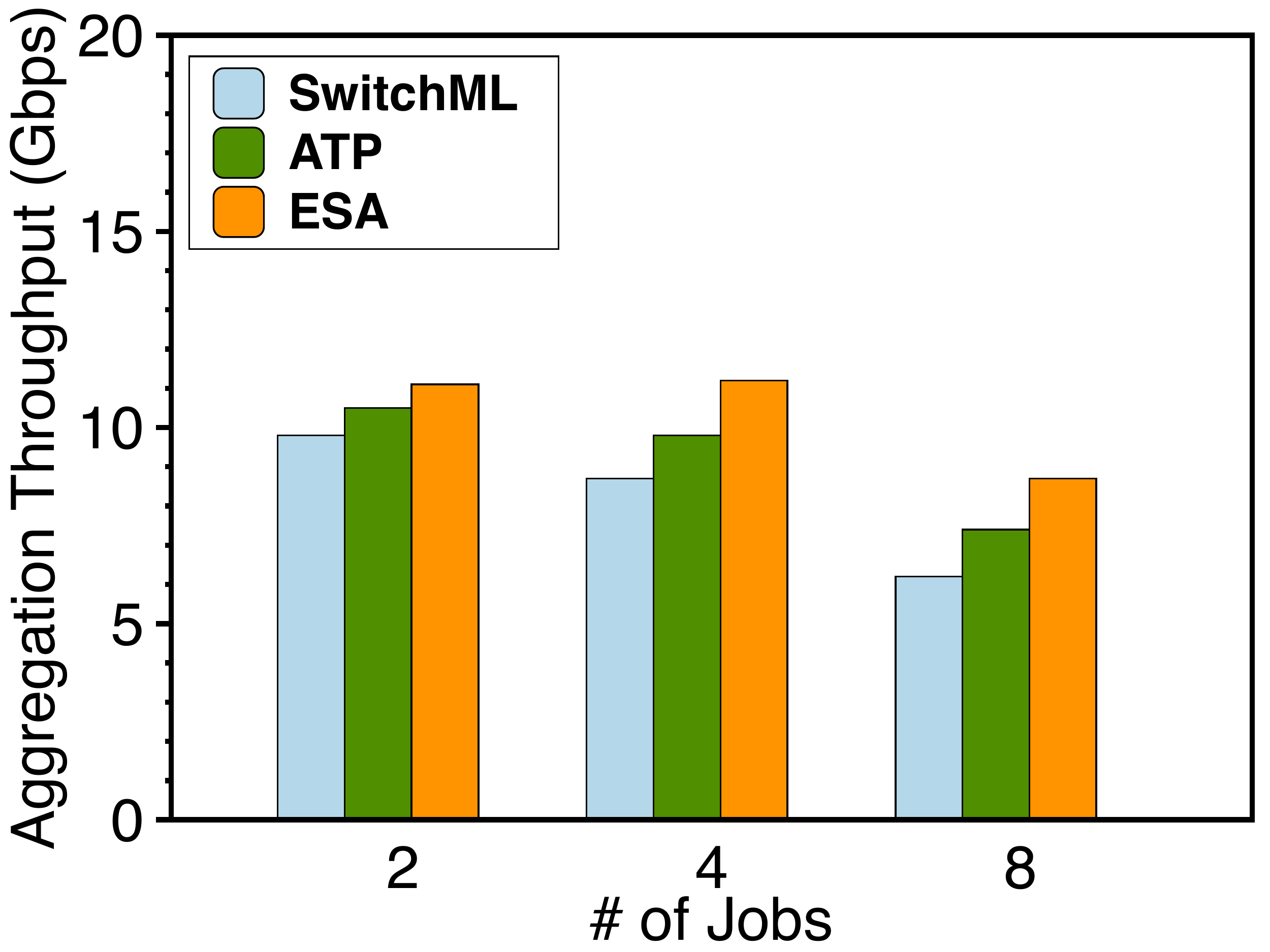}
    }
    \vspace{-0.1in}
    \caption{Aggregation Throughout.}
    \vspace{-0.1in}
    \label{fig:testbed:micro}
\end{figure}

\vspace{-0.05in}
\subsubsection{Microbenchmark} We illustrate the aggregation efficiency by the metric aggregation throughput, which is the volume of parameters (Byte) each worker received per second. We conduct two settings, the first is to fix the number of jobs to 4 and change the tensor size, the second is to fix the tensor size to 4MB and change the number of jobs. We use the first 8 containers \{$w_1\sim w_8$\} to be the workers and the remaining 2 to be the PSes \{$p_1, p_2$\}. All jobs with the odd ID are located on \{$w_1\sim w_4$, $p_1$\}, other jobs are on \{$w_5\sim w_8$, $p_2$\}. Figure~\ref{fig:testbed:micro} shows the results. \sys outperforms SwitchML and ATP by up to 1.39$\times$ and 1.18$\times$, respectively. We also observe that the speedup of INAs is improved with larger tensor size and less current jobs. 

\begin{figure*}[htb!]
    \captionsetup{skip=0pt}
    \centering
    \subfigure[DNN A]{
	\includegraphics[width=0.3\linewidth]{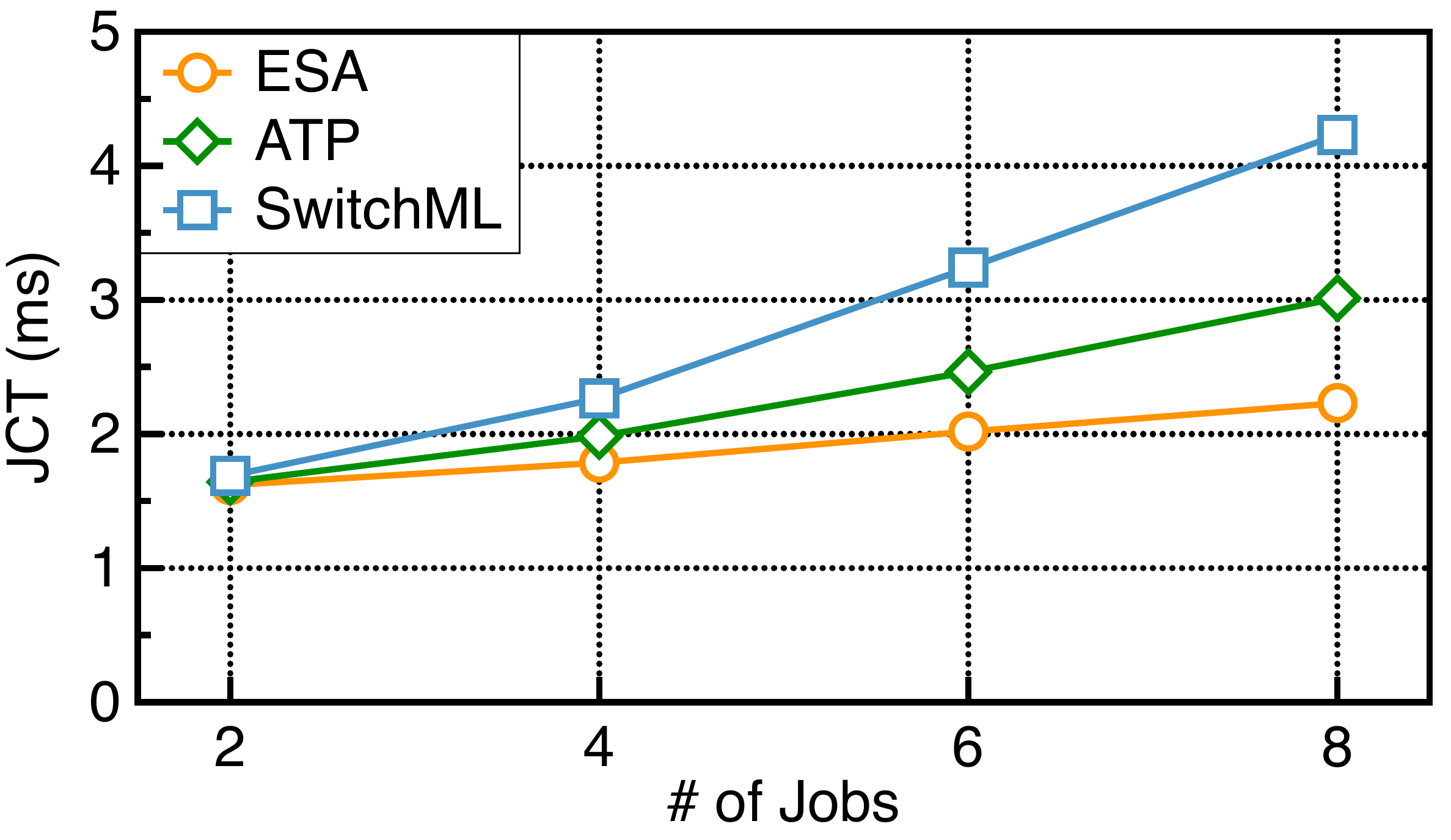}
    }
    \subfigure[DNN B]{
        \includegraphics[width=0.3\linewidth]{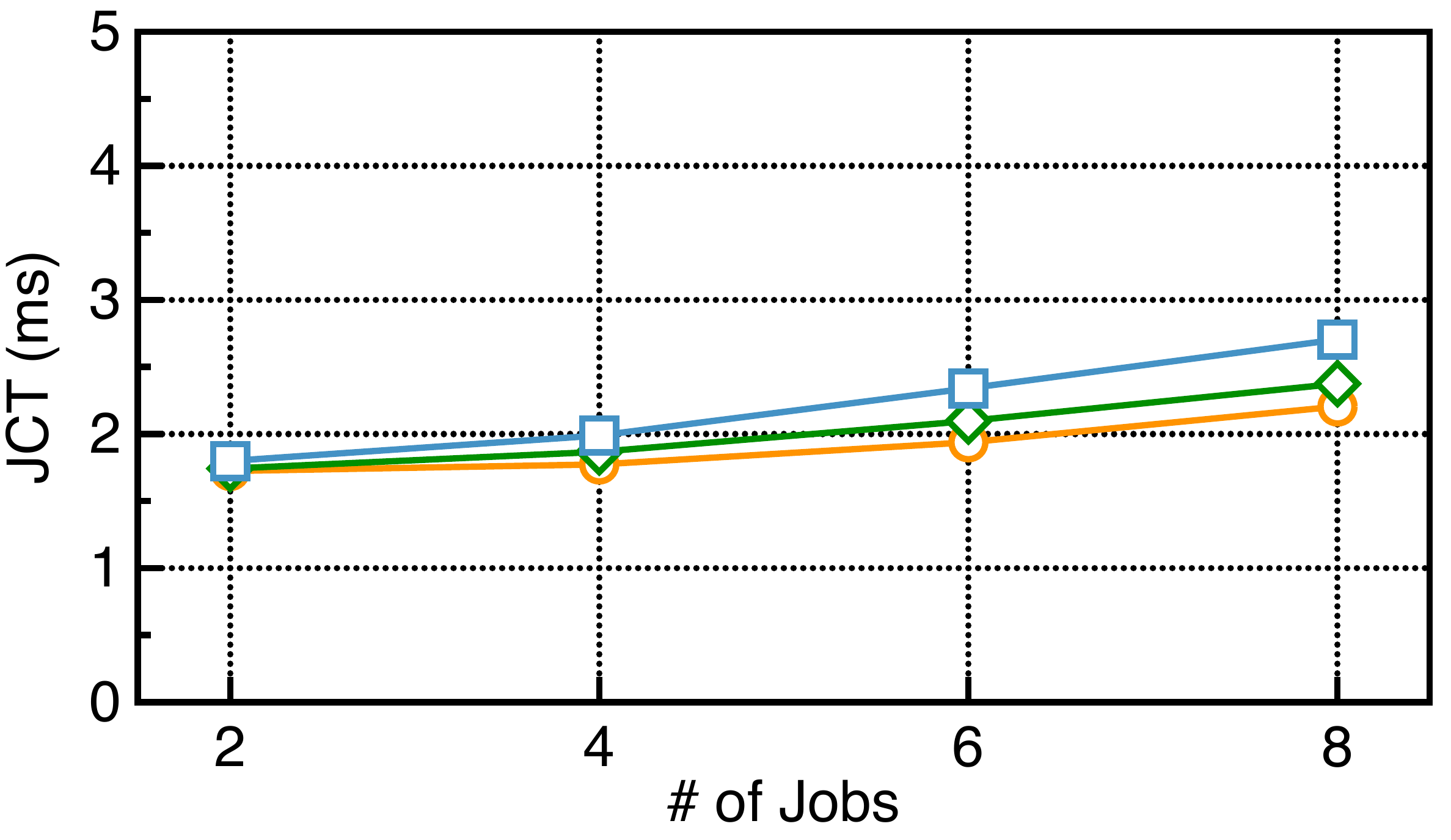}
    }
    \subfigure[A:B = 1:1]{
	\includegraphics[width=0.3\linewidth]{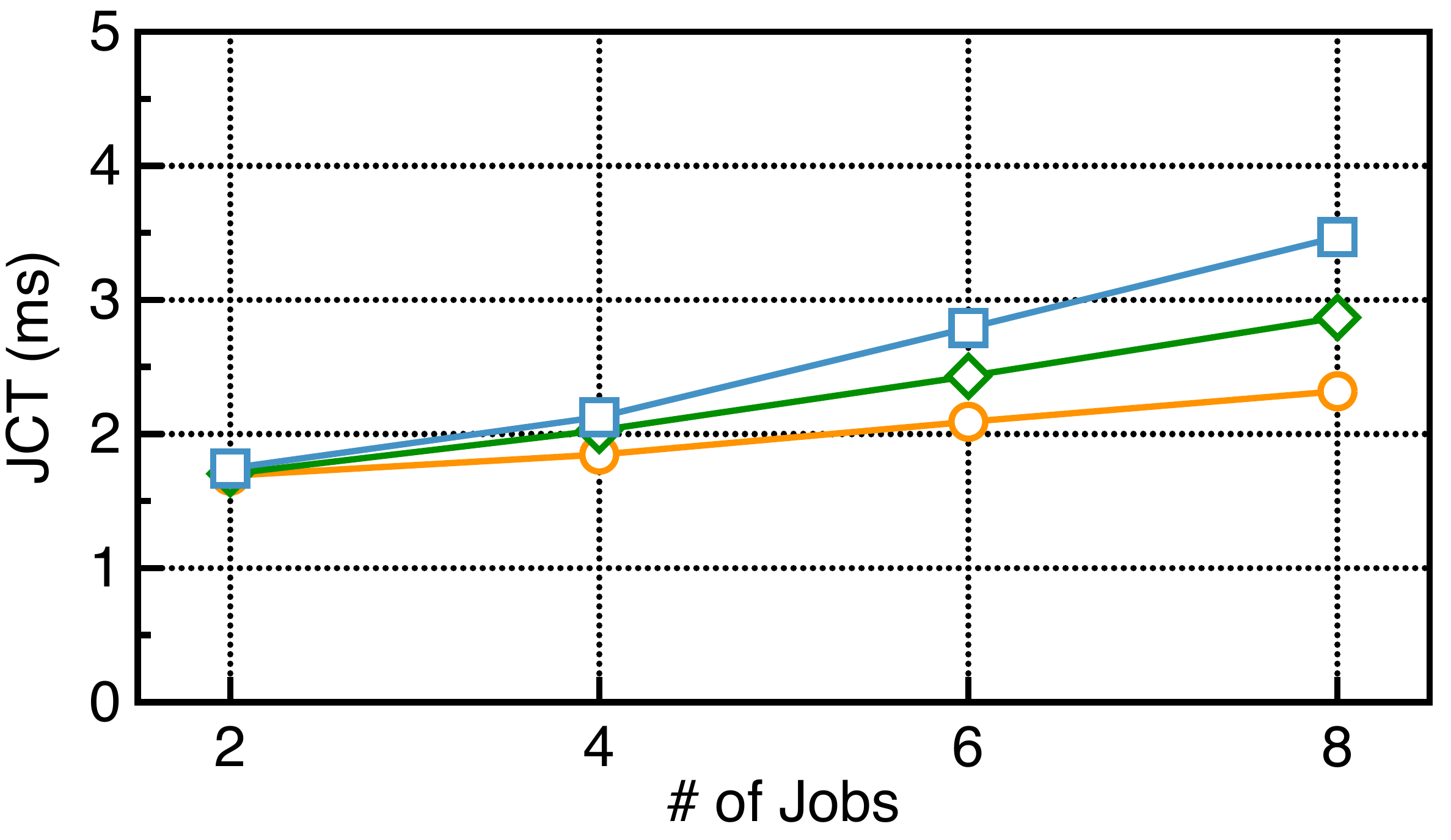}
    }
    \caption{Average JCT with different number of jobs.}
    \vspace{-0.15in}
    \label{figure:sim1}
\end{figure*}

\begin{figure*}[htb!]
    \captionsetup{skip=0pt}
    \centering
    \subfigure[DNN A]{
	\includegraphics[width=0.3\linewidth]{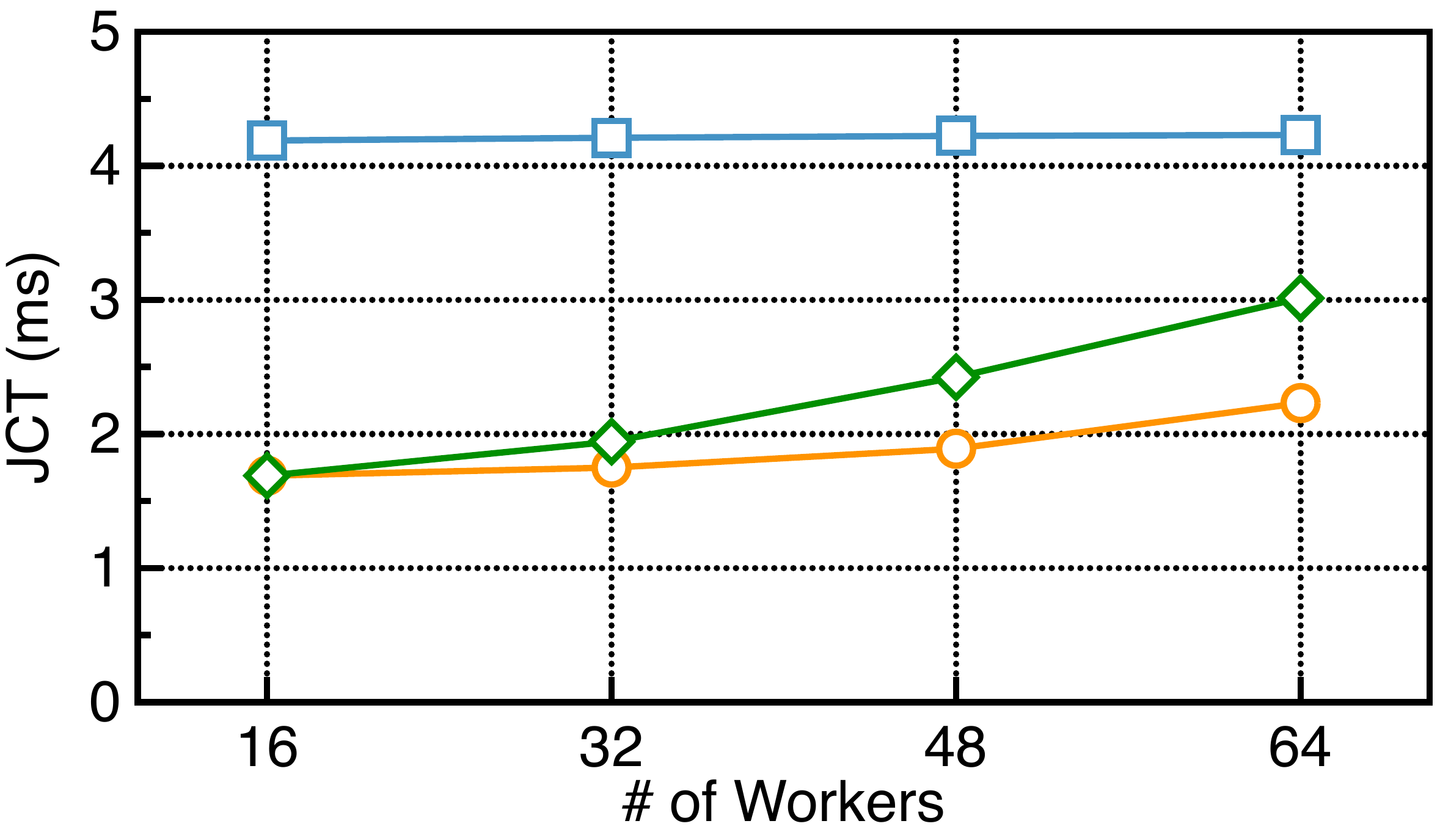}
    }
    \subfigure[DNN B]{
        \includegraphics[width=0.3\linewidth]{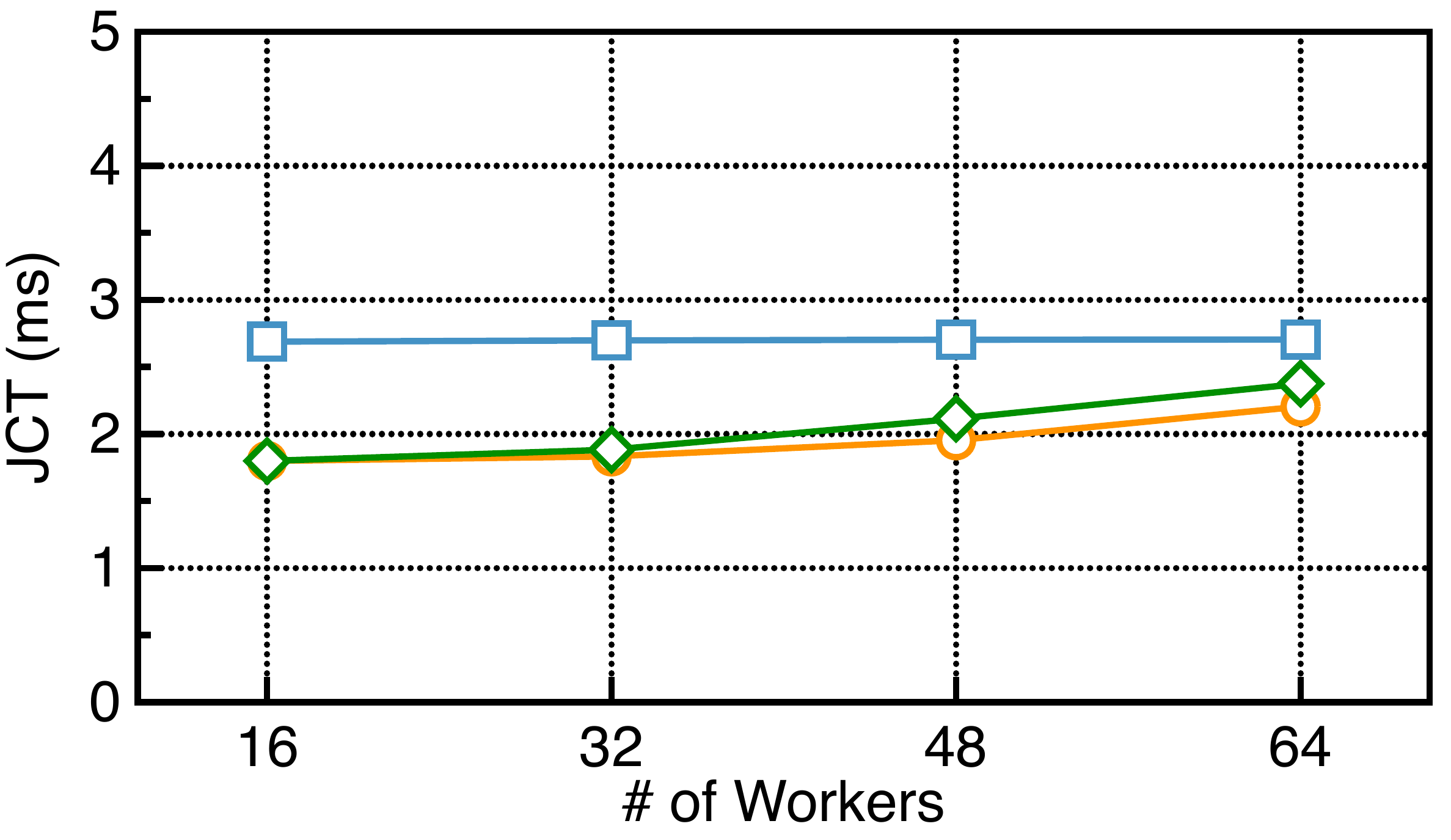}
    }
    \subfigure[A:B = 1:1]{
	\includegraphics[width=0.3\linewidth]{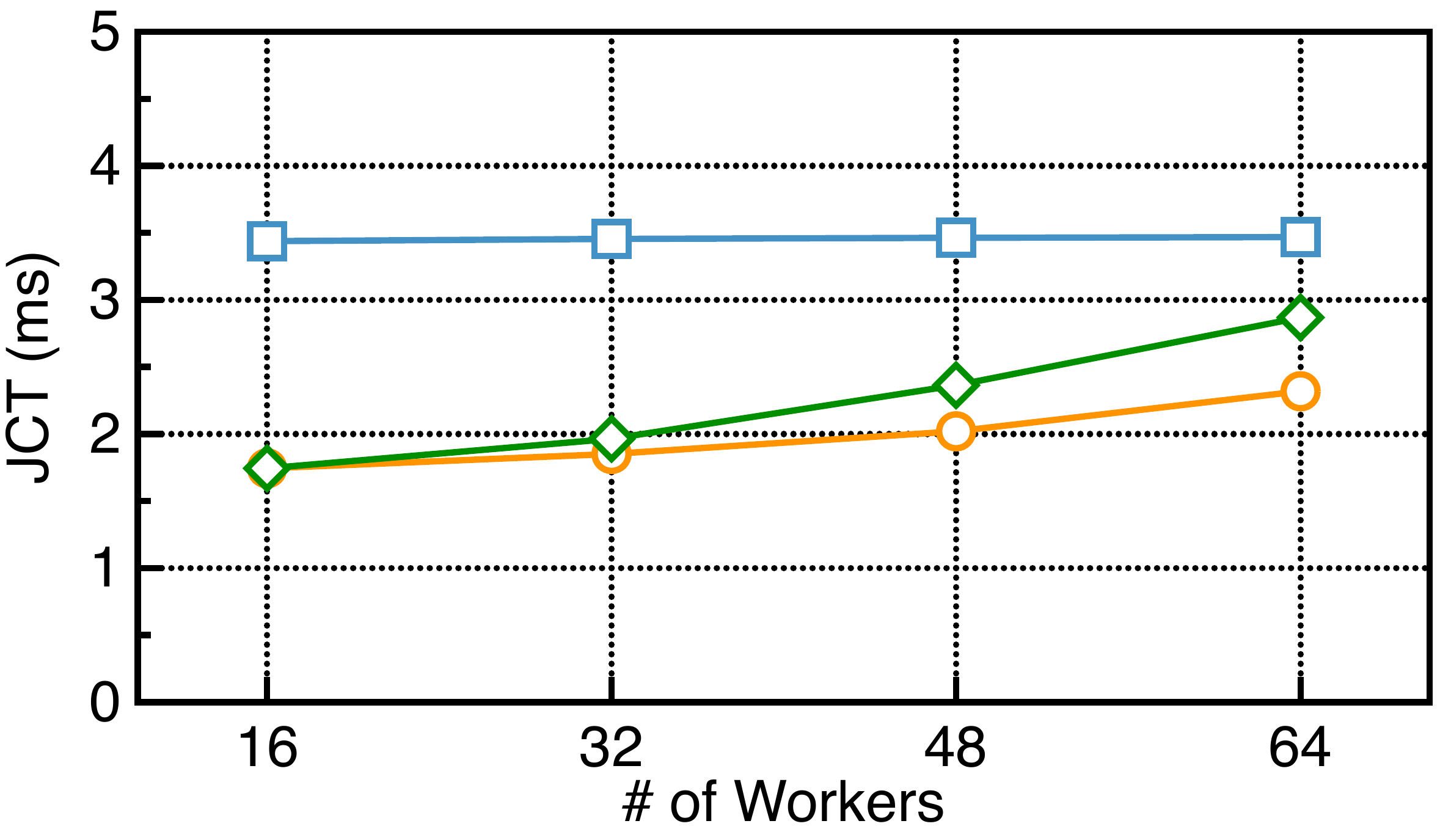}
    }
    \caption{Average JCT with different number of workers.}
    \label{figure:sim2}
    \vspace{-0.1in}
\end{figure*}

\vspace{-0.1in}
\subsection{Simulations with more Workers and Jobs}
\label{sec:eval:sim}
\vspace{-0.05in}

Considering the limited scale of our testbed, here we evaluate \sys under more jobs and larger topologies with NS3~\cite{ns3} simulation as a supplement.
\vspace{-0.15in}
\subsubsection{Simulation Setup}
\vspace{-0.07in}

\parab{Topology:} We set up a single switch topology, one switch with 64 100Gbps links to 64 servers. The base RTT is $10\mu s$.  The switch memory reserved for INA is 5MB. We use the same packet size (306Byte) as the setting of ATP~\cite{atp}.



\parab{Simulation Model:} Our simulation considers the communication and computation overlapping~\cite{poseidon} of DNN training. For simplicity, we assume that all DNNs have only two layers with the same size, and each layer is evenly divided into two tensor partitions~\cite{bytescheduler}. Since the backward propagation (BP) starts from the back layer to compute the gradients, we assume that the first tensor partition of the second layer is transmitted first, followed by the first layer, and finally the second tensor partition of the second layer. When a worker receives all aggregation results of the first layer, it can directly start the computation of the first layer. While the computation of the second layer must wait for the computation completion of the first layer and the arrival of the aggregation results.

\parab{Workloads:} We assume two types of DNNs, the first one (A) is communication-intensive with a tensor partition size of 4MB and a computation time of $0.32ms$ per layer (theoretical communication time: computation time = 2:1). The second one (B) is computation-intensive with a tensor partition size of 2MB and a computation time of $0.64ms$ per layer (theoretical communication time: computation time = 1:2).

\parab{Baselines and Metrics:} We compare \sys with two state-of-the-art INAs: ATP and SwitchML. For SwitchML, each job shares the memory equally. We measure the average job completion time (JCT), which is the average of the computation completion time minus the communication start time of the previous iteration for all  jobs.

\parab{Parameter Settings:}
1) Job placement: we assume no overlapping usage of servers among different jobs and each server contains at most one worker.
2) Job start time: to reflect the real situation, we need to avoid every DNN job starting exactly at the same time, here we use a random variable $t$ as the job start time, $t\sim U(0,1ms)$.
3) Computation speed variance: Considering the different computation speeds of different workers, we add a jitter on the sending side, whose maximum value is 300us, i.e., $jitter\sim U(0,300us)$, which is approximate to the computation time of a tensor partition.
4) Parameter server: For ATP and \sys, a parameter server is required for each job. Without loss of generality, we add 8 extra servers as the PSes for the 8 jobs and 64 workers case.
5) Priority setting: we calculate the priority according to the formula of $\S$\ref{sec:design:priority}. Since our DNNs have two layers, all $L_j$ is 2. The $l$ can be 1 or 2 according to which layer. For DNN A, the $\frac{Comm_j}{Comp_j}=2$, for DNN B, the $\frac{Comm_j}{Comp_j}=0.5$. We use the remaining communication plus the computation time to estimate the $T_j$.

\vspace{-0.15in}
\subsubsection{Simulation Results}
\vspace{-0.1in}
%
%

We have two groups of simulations. In the first group, we fix the worker number in each job to be 8, and change the number of jobs. In the second group, we fix the number of jobs to be 8, and change the number of  workers in each job (For simplicity, we assume the number of workers in each job is the same). We conduct three simulations in each group: All jobs are DNN A; All jobs are B; The ratio of A to B is 1:1.

\parab{Speedup with different number of jobs.} Figure~\ref{figure:sim1} shows the JCT under different number of jobs. \sys outperforms SwitchML and ATP by up to 1.89$\times$ and 1.35$\times$. We also observe that the speedup of \sys becomes more significant with more jobs. The reason is that with more jobs, the contention of switch will be more severe. This is consistent with our testbed.

\parab{Speedup with different number of workers.} Figure~\ref{figure:sim2} shows the JCT for different INA solutions under different number of workers. Y-axis indicates the JCT, and X-axis indicates the number of workers in each job. \sys outperforms SwitchML and ATP under all cases. \sys obtains more improvement over ATP with more workers. This is expected because with more workers, the synchronization cost of aggregation will increase, thus preemption can obtain more benefit.
\vspace{-0.1in}
\begin{figure}[htb!]
    \centering
    \subfigure[DNN A]{
        \includegraphics[width=0.472\linewidth]{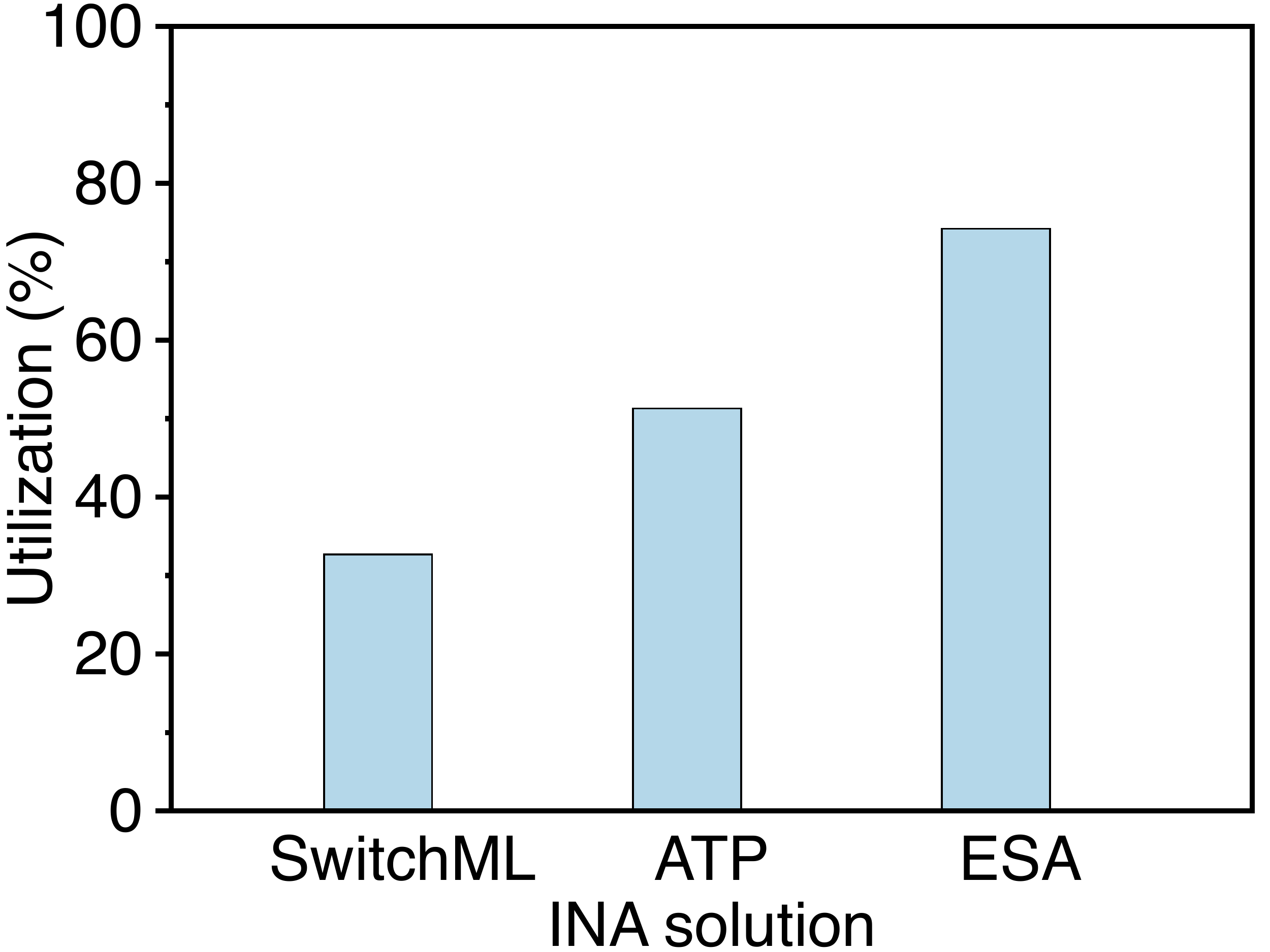}
    }
    \subfigure[DNN B]{
	\includegraphics[width=0.472\linewidth]{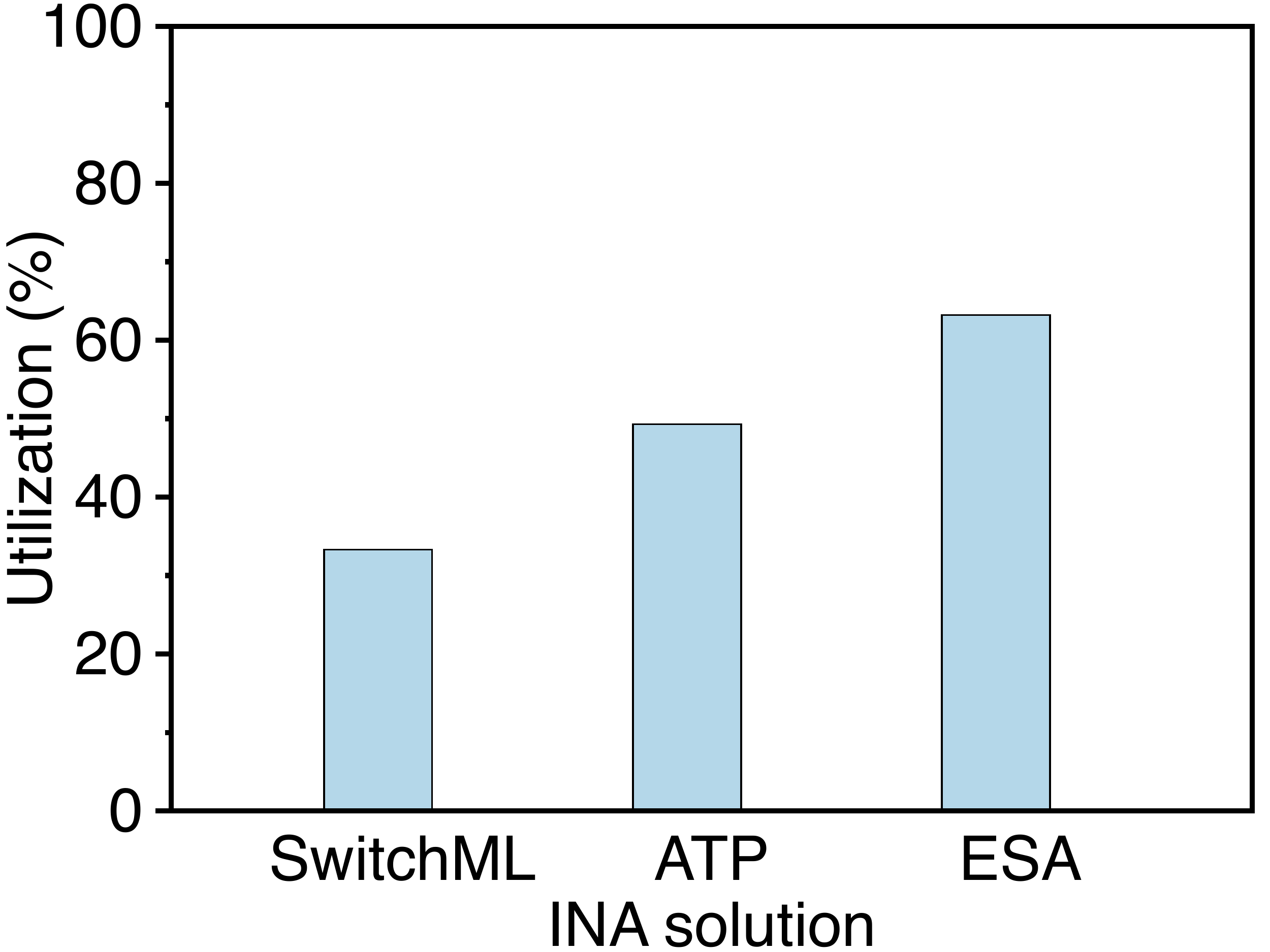}
    }
    \vspace{-0.1in}
    \caption{Measurement on switch memory utilization.}
    \vspace{-0.15in}
    \label{figure:dive1}
\end{figure}

\begin{figure}[htb!]
    \centering
    \subfigure[DNN A]{
        \includegraphics[width=0.472\linewidth]{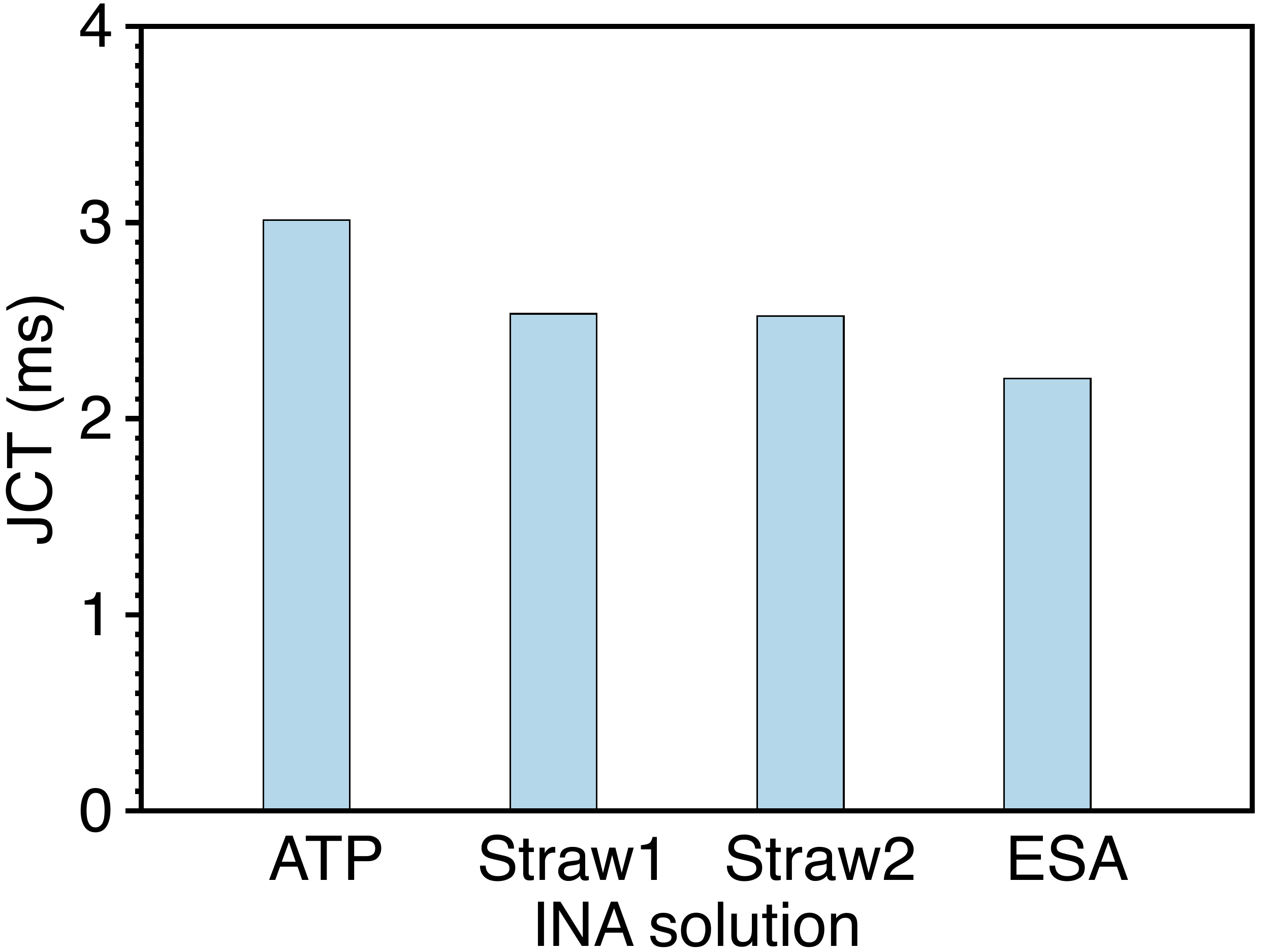}
    }
    \subfigure[A:B = 1:1]{
	\includegraphics[width=0.472\linewidth]{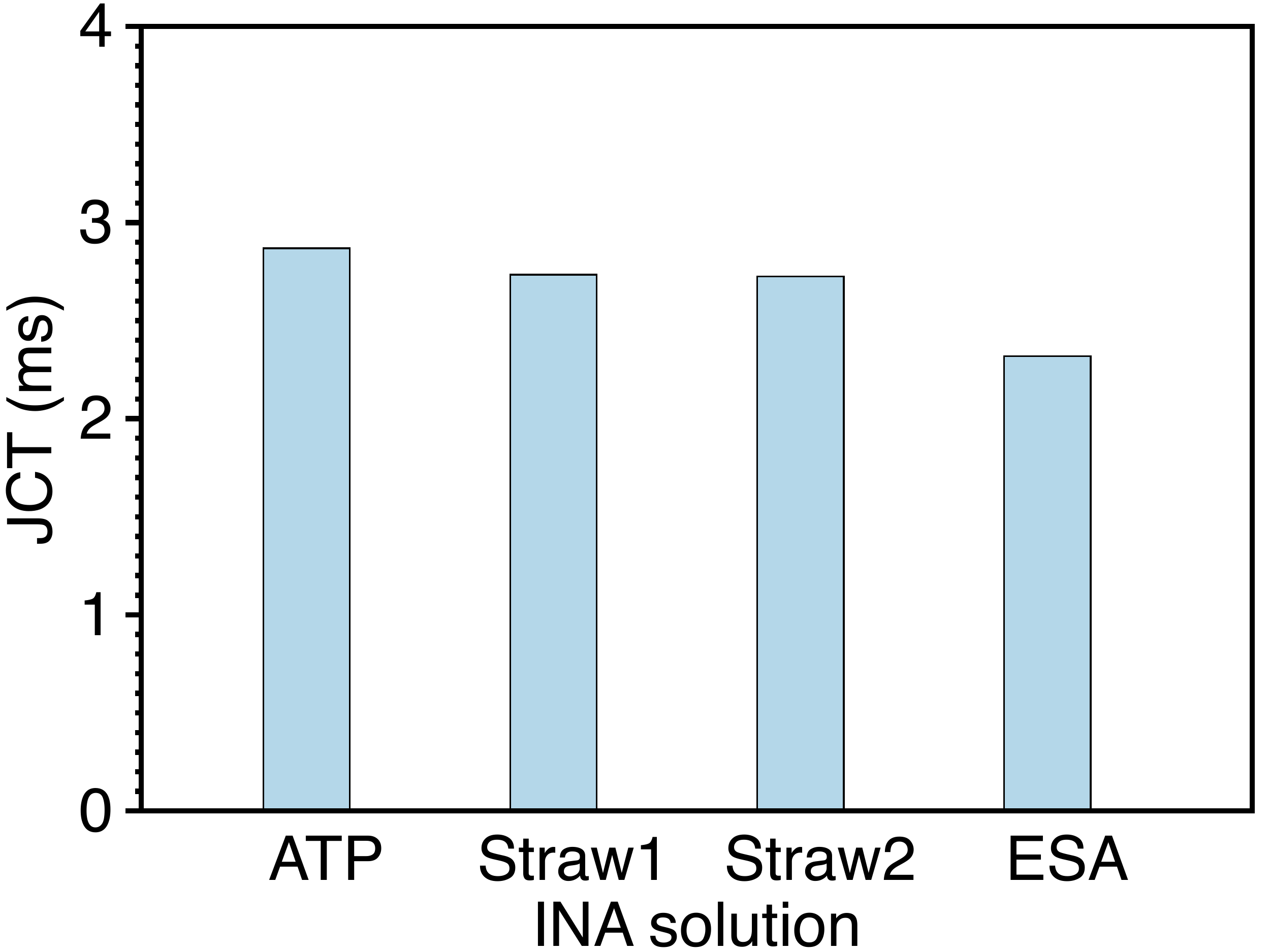}
    }
    \vspace{-0.15in}
    \caption{The speedup of priority scheduling.}
    \vspace{-0.15in}
    \label{figure:dive2}
\end{figure}

\vspace{-0.05in}
\subsection{Deep Dive}
\label{sec:eval:deepdive}
\vspace{-0.1in}

\parab{The improvement on switch memory utilization:}
Here we explore the improvement of enabling preemption allocation on switch memory utilization. First we give the metric definition for utilization. Previous INA work~\cite{atp} uses aggregation throughput to measure the aggregation efficiency. This value has an upper bound, which is the volume of all gradients divided by the bandwidth (100Gbps). We define the switch memory utilization as the aggregation throughput divided by the upper bound. For the multi-job scenario, we calculate the average utilization of each job.
Figure~\ref{figure:dive1} shows switch memory utilization measured on two types of DNNs. We simulate 8 jobs, each with 8 workers, other settings are the same with $\S$\ref{sec:eval:sim}. \sys outperforms SwitchML and ATP by 2.27$\times$ and 1.45$\times$ for DNN A. The numbers for DNN B are 1.9$\times$ and 1.28$\times$. We observe more improvement on communication-intensive models, i.e, DNN A, which consists with our testbed.

\parab{The speedup of priority scheduling:} Our design essentially has two components: preemptive allocation and priority scheduling. Here we measure the improvement of priority scheduling. We make two straw-man preemption INA solutions: the first one always does preemption upon hash conflict; the second one has a 50-50 chance to do preemption. We compare \sys with these two solutions and ATP by simulation. There are 8 jobs each with 8 workers. Other settings are in $\S$\ref{sec:eval:sim}. Figure~\ref{figure:dive2} shows the average JCT under two settings 1) all jobs are DNN A; 2) DNN A and DNN B each has 4 jobs.
\sys, Straw1 and Straw2 outperforms ATP by 1.35$\times$, 1.19$\times$ and 1.19$\times$ for DNN A. For the mixed models setting, the numbers are 1.22$\times$, 1.05$\times$ and 1.05$\times$. \sys performs better than the Straw-man solutions and shows more improvement with a mixed models setting. These prove that \sys benefits from the priority scheduling.



    \vspace{-0.15in}
\section{Related Work}
\vspace{-0.15in}
\parab{Communication Optimization in DLT.}
Besides the closely related INA discussed in the $\S$\ref{sec:background}, many other solutions are proposed to optimize the DLT communication~\cite{wan2020rat, wang2020divide, xia2019rethinking, wang2020domain}.
For example, from the application level, gradient sparsification and quantization reduce the communication volume in one iteration. Large batch training and periodic communication reduce the communication rounds. From the framework level, works like BlueConnect~\cite{blueconnect} and PLink~\cite{plink} design a network topology-aware framework to optimize the communication pattern. SSP relaxes synchronization requirements to minimize the synchronization cost. Poseidon overlaps the communication with the computation and ByteScheduler further designs a generic framework with the tensor partitioning and scheduling.
Other works also explore the optimization for model parallelism, such as GPipe~\cite{gpipe} and PipeDream~\cite{pipedream}.



\parab{Resource Management in Deep Learning Clusters.} Besides the switch memory, other resources like GPUs and network bandwidth also need carefully scheduling to guarantee resource efficiency and job performance~\cite{zhao2015rapier, li2017rate, bai2015information, susanto2016stream, zhang2016coda, chowdhury2014efficient, chen2018auto}. Gandiva~\cite{gandiva} manage the resource in GPU clusters via time-sharing and Migration. Optimus~\cite{optimus} is an online resource scheduler which is further optimized for the dynamically workload by building the resource-performance model. Tiresias~\cite{tiresias} obverses that a DLT job's execution time is often ignostic and proposes the \textit{Discretized TwoDimensional Gittins index} and \textit{Discretized Two-Dimensional LAS} to minimize the average JCT.

    \vspace{-0.17in}
\section{Conclusion}
\vspace{-0.12in}
This paper introduces the preemptive allocation and priority scheduling of switch memory resource for INA. Although existing INA implementations have nearly exhausted the resources of programmable switches, \sys manages to implement the priority preemption mechanism on the data-plane and guarantee the reliability of aggregations. We evaluate \sys on our V100 testbed together with the NS3 simulation. Experiment results show that \sys can outperform the state-of-the-art INAs by up to 1.35$\times$ in terms of average JCT.
	\bibliographystyle{plain}
	\bibliography{ref}

\end{document}